\documentclass[
    reprint,
	amsmath,
    amssymb,
	aps,
	prl,
    floatfix,
    superscriptaddress
]{revtex4-1}

\usepackage[english]{babel}
\usepackage[utf8x]{inputenc}
\usepackage[T1]{fontenc}
\usepackage{braket}
\usepackage{mathtools}
\usepackage{flushend}

\usepackage{amsmath}
\usepackage{graphicx}
\usepackage{dcolumn}
\usepackage{bm}
\usepackage{array}
\usepackage[shortlabels]{enumitem}
\newcolumntype{?}{!{\vrule width 1pt}}
\usepackage[colorinlistoftodos]{todonotes}
\usepackage{xr-hyper}
\usepackage{xcolor}
\usepackage{hyperref}
\hypersetup{colorlinks,allcolors=blue}

\usepackage{bibunits}
\defaultbibliographystyle{apsrev4-1}
\usepackage{etoolbox}
\usepackage{multirow}
\usepackage{booktabs, makecell}
\usepackage{threeparttable}

\makeatletter
\newcommand*{\newbibstartnumber}[1]{%
  \apptocmd{\thebibliography}{%
    \global\c@NAT@ctr #1\relax
    \addtocounter{NAT@ctr}{-1}%
  }{}{}%
}
\makeatother

\makeatletter
\let\cat@comma@active\@empty
\makeatother

\definecolor{revision2}{RGB}{0,0,0}
\definecolor{revision}{RGB}{0,0,0}
\begin{document}

\title{Direct prediction of phonon density of states with Euclidean neural networks}
\author{Zhantao Chen}
\thanks{These authors contributed equally to this work.}
\affiliation{Quantum Matter Group, MIT, Cambridge, MA 02139}
\affiliation{Department of Mechanical Engineering, MIT, Cambridge, MA 02139}

\author{Nina Andrejevic}
\thanks{These authors contributed equally to this work.}
\affiliation{Quantum Matter Group, MIT, Cambridge, MA 02139}
\affiliation{Department of Materials Science and Engineering, MIT, Cambridge, MA 02139}

\author{Tess Smidt}
\thanks{These authors contributed equally to this work.}
\affiliation{Computational Research Division, Lawrence Berkeley National Laboratory, Berkeley, CA 94720}
\affiliation{Center for Advanced Mathematics for Energy Research Applications, Lawrence Berkeley National Laboratory, Berkeley, CA 94720}

\author{Zhiwei Ding}
\affiliation{Department of Materials Science and Engineering, MIT, Cambridge, MA 02139}

\author{Qian Xu}
\affiliation{Department of Mechanical Engineering, MIT, Cambridge, MA 02139}

\author{Yen-Ting Chi}
\affiliation{Department of Materials Science and Engineering, MIT, Cambridge, MA 02139}

\author{Quynh T. Nguyen}
\affiliation{Department of Physics, MIT, Cambridge, MA 02139}

\author{Ahmet Alatas}
\affiliation{Advanced Photon Source, Argonne National Laboratory, Lemont IL 60439}

\author{Jing Kong}
\affiliation{Department of Electrical Engineering and Computer Science, MIT, Cambridge, MA 02139}
\author{Mingda Li}
\thanks{Corresponding author.\\\href{mailto:mingda@mit.edu}{mingda@mit.edu} \vspace{0.5cm}}
\affiliation{Quantum Matter Group, MIT, Cambridge, MA 02139}
\affiliation{Department of Nuclear Science and Engineering, MIT, Cambridge, MA 02139}

\date{\today}

\begin{abstract}
	Machine learning has demonstrated great power in materials design, discovery, and property prediction. However, despite the success of machine learning in predicting discrete properties, challenges remain for continuous property prediction. The challenge is aggravated in crystalline solids due to crystallographic symmetry considerations and data scarcity. Here we demonstrate the direct prediction of phonon density of states using only atomic species and positions as input. We apply Euclidean neural networks, which by construction are equivariant to 3D rotations, translations, and inversion and thereby capture full crystal symmetry, and achieve high-quality prediction using a small training set of $\sim10^{3}$ examples with over 64 atom types. Our predictive model reproduces key features of experimental data and even generalizes to  materials with unseen elements, {\color{revision}and is naturally suited to efficiently predict alloy systems without additional computational cost}. We demonstrate the potential of our network by predicting a broad number of high phononic specific heat capacity materials. Our work indicates an efficient approach to explore materials’ phonon structure, and can further enable rapid screening for high-performance thermal storage materials and phonon-mediated superconductors.
\end{abstract}

\maketitle

\section{Introduction}

One central objective of materials science is to establish structure-property relationships; that is, how specific atomic arrangements lead to certain macroscopic functionalities. This question is historically addressed through trial-and-error of a combination of structure and property characterization, theory, and calculation. However, recent advances in machine learning (ML) suggest a paradigm shift in how structure-property relationships can be directly constructed \cite{ramprasad2017machine,schmidt2019recent}. To date, ML has seen success in a growing spectrum of materials applications, including materials discovery and design \cite{raccuglia2016machine,oliynyk2016high,gomez2018automatic,LIU2017discovery_design}, process automation and optimization \cite{Alan2018Chimera,Granda2018}, and prediction of materials' mechanical (elastic moduli) \cite{pilania2013accelerating, isayev2017universal, xie2018crystal, chen2019graph}, thermodynamic and thermal transport (formation enthalpy, thermal conductivity, Debye temperature, heat capacity) \cite{carrete2014finding,van2016high,isayev2017universal,chen2019graph,tawfik2020predicting,mortazavi2020machine}, and electronic (bandgap, superconductivity, topology) properties \cite{ward2016general,zhuo2018predicting,xie2018crystal,dong2019bandgap,meredig2018can,stanev2018machine,andrejevic2020machine,claussen2020detection,scheurer2020unsupervised}, and atomistic potentials (potential energy surfaces and force constants) \cite{li2015molecular,chmiela2017machine,kruglov2017energy,glielmo2017accurate,botu2017machine,legrain2018vibrational,zhang2019active}. Most property prediction studies consider a low-dimensional output consisting of one or few discrete points. However, the prediction of continuous properties from limited input information remains challenging due to the output complexity and finite data volume. Moreover, for crystalline solids, the crystallographic symmetry poses additional constraints on a generic neural network.


\begin{figure*}
	\centering
	\includegraphics[width=0.9\linewidth]{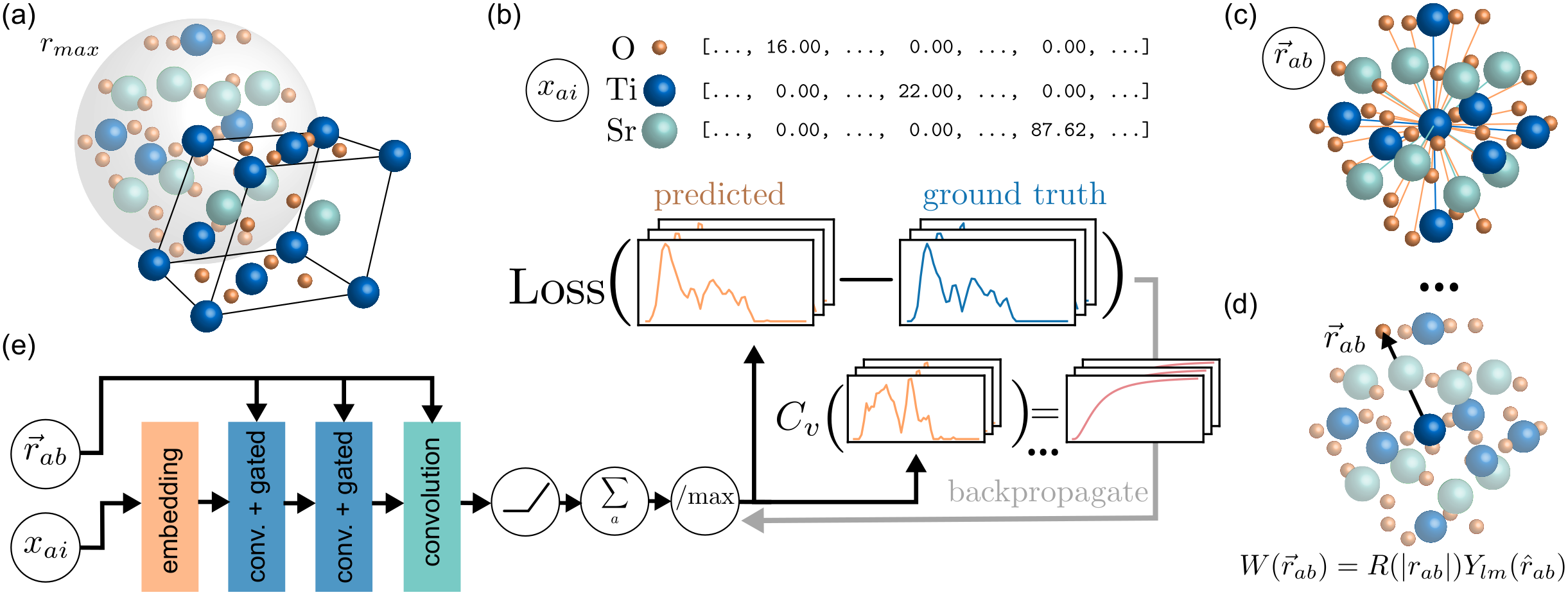}
	\caption{Overview of the E(3)NN architecture for phonon DoS prediction. (a) Crystals are converted to periodic graphs by considering all periodic neighbors within a radial cutoff $r_{max} = 5 \text{\normalfont\AA}$. The example of SrTiO\textsubscript{3} is shown. (b) Atom types are encoded as a mass-weighted one-hot encoding. (c) Edges join neighboring atoms and store the relative distance vector from the central atom to neighbor. (d) The radial distance vectors are used for the continuous convolutional filters $W(\vec{r}_{ab})$ comprising learned radial functions and spherical harmonics. (e) The E(3)NN operates on the node and edge features using convolution and gated nonlinear layers. The result is passed to a final activation, aggregation, and normalization to generate the predicted output. The network weights are trained by minimizing the loss function between the predicted and ground-truth phonon DoS.}
	\label{fig:networkworkflow}
\end{figure*}

In this work, we build a ML-based predictive model that directly outputs the phonon density of states (DoS) using atomic structures as input. Phonon DoS is a key determinant of materials' specific heat and vibrational entropy and plays a crucial role in interfacial thermal resistance \cite{Wu2019interfacial}. It is also tightly linked to thermal and electrical transport \cite{Giustino2017elph} and  superconductivity \cite{bardeen1957theory}. However, the acquisition of experimental and computed phonon DoS is nontrivial due to limited inelastic scattering facility resources and high computational cost of \textit{ab initio} calculations for complex materials \cite{baroni2001phonons,Giustino2017elph}. \textcolor{revision}{Moreover, the phonon calculations in alloys systems pose significant challenge. Some existing approaches like virtual crystal approximation (VCA) can fail both qualitatively and quantitatively without well-controlled approximations \cite{Seyf2017Ase}.} This calls for an approach that acquires phonon DoS more efficiently, \textcolor{revision}{especially for alloy systems}. To build such a model, we employ a Euclidean neural network (E(3)NN) which naturally operates on 3D geometry and is equivariant to 3D translations, rotations, and inversion  \cite{thomas2018tensor,kondor2018clebsch,3dsteerable, e3nn_2020_3724963}. E(3)NNs preserve all geometric information of the input and eliminate the need for expensive (approximately 500 fold) data augmentation. Additionally, all crystallographic symmetries of input data are preserved by the network \cite{Tess2020Euclidean}. In this work, we use E(3)NNs as implemented in the open-source \texttt{e3nn} repository \cite{e3nn_2020_3724963} which merges implementations of Ref.~\cite{thomas2018tensor} and Ref.~\cite{3dsteerable} and additionally implements inversion symmetry. High-fidelity phonon DoS predictions are achieved using the density functional perturbation theory (DFPT)-based phonon database \cite{petretto2018high} containing phonon DoS data of approximately 1,500 crystalline solids. Our predictive model can capture the main features of phonon DoS, even for crystalline solids with unseen elements. By predicting the phonon DoS in 4,346 new crystal structures, we identify a list of high heat capacity materials, supported by additional DFPT calculations. Our work offers an efficient technique to acquire phonon DoS directly from atomic structure, making it suitable for high throughput materials design with desirable phonon-related properties.

\section{Euclidean neural network for phonon DoS prediction}

Crystal structures operated on by the E(3)NN are first converted into a periodic graphs where atoms are nodes $N$ with edges $E$ connecting neighboring atoms within a specified radial cutoff, including periodic images (Figure \ref{fig:networkworkflow}). Each edge $e_{ab} \in E$ stores the radial distance vector between atom $a$ and neighbor $b$, $\vec{r}_{ab}$, up to some radial cutoff $|r_{max}|$, and is used by the convolutional kernels of the E(3)NN. The input node features are scalars that captures its atomic type and mass using one-hot encoding; for instance, a hydrogen atom is encoded as $x_{\mathrm{H}}=[m_{\mathrm{H}},0,\ldots,0]$. After an initial embedding layer which takes the 118-length one-hot mass-weighted encodings to 64 scalar features, the constructed graph is then passed to the E(3)NN, which iteratively operates on the features with multiple ``Convolution and Gated Block'' layers as described for the L1Net of Ref.\cite{e3nnQM9} (see \hyperref[sec:SIE(3)NNdetail]{Supplementary Material} for more details). After the final layer, which consists of only a convolution, all resulting node features are summed and passed through a final activation (ReLU) and normalization layer to predict the phonon DoS, comprising 51 scalars. The absolute magnitude of the phonon DoS can easily be recovered from the normalized DoS by noticing that $\int g(\omega)\mathrm{d}\omega = 3N$, where $N$ is the number of atoms in the unit cell; thus, we ensure that normalization of the DoS does not compromise meaningful prediction. The E(3)NN weights are optimized by minimizing the mean squared error (MSE) loss function between the DFPT-computed DoS $\bm{g}$ and E(3)NN-predicted $\hat{\bm{g}}$. The full network structure is provided in the \hyperref[sec:SIE(3)NNdetail]{Supplementary Material}. 

\begin{figure*}[t]
	\centering
	\includegraphics[width=0.9\linewidth]{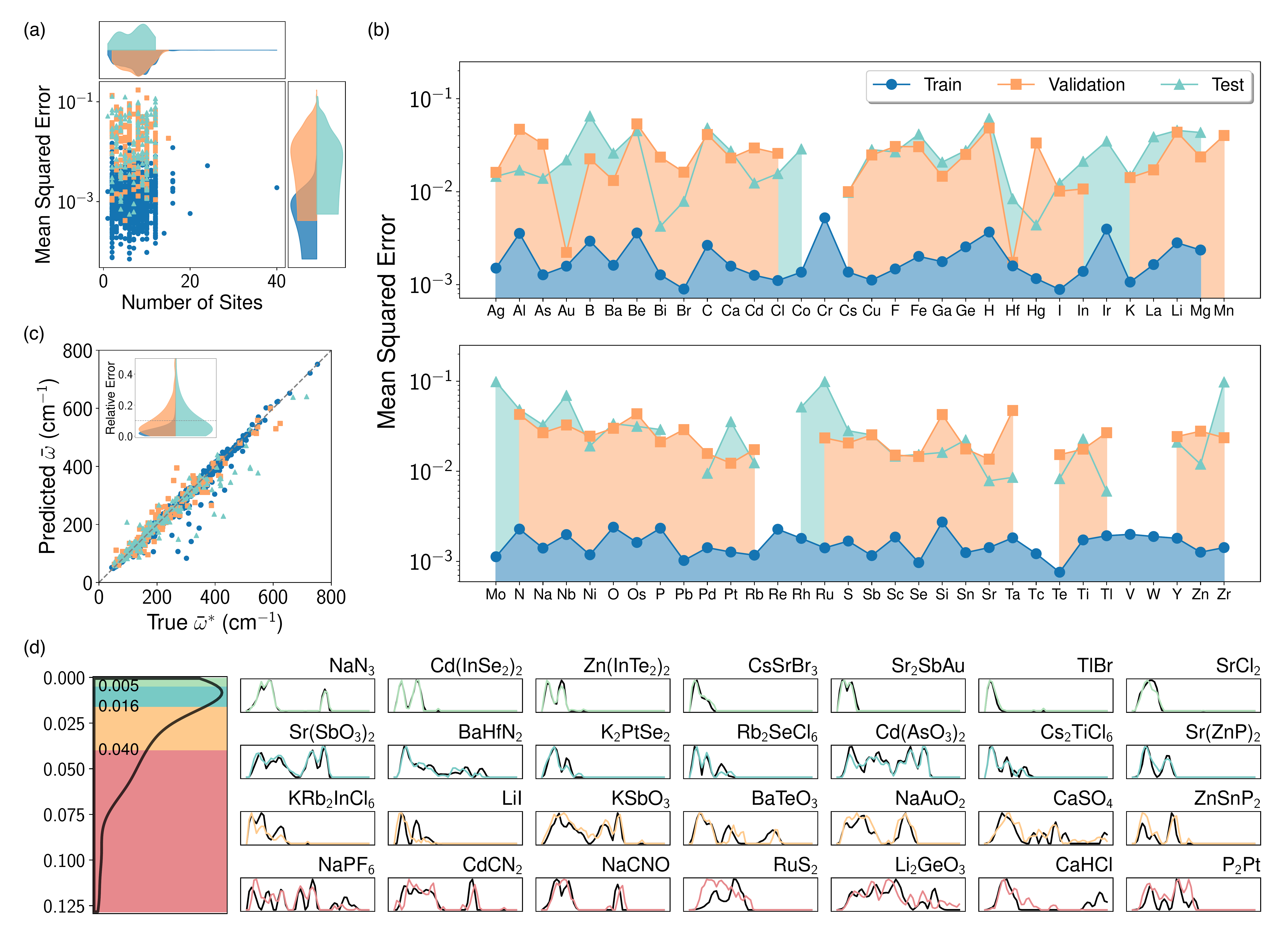}
	\caption{Performance of the Euclidean neural network-based predictive model. (a) Mean squared error versus total number of sites in a unit cell in training (blue), validation (orange), and test (green) sets. (b) Average mean squared error of compounds containing each element. (c) Comparison between E(3)NN-predicted average phonon frequency and ground truth. The inset shows the relative error $|\bar{\omega}-\bar{\omega}^{\ast}|/\bar{\omega}^{\ast}$ distribution of the three datasets. (d) Randomly selected examples in the test set within each error quartile. (Left) MSE distribution showing that it is heavily peaked in the 1$^{\text{st}}$ and 2$^{\text{nd}}$ quartiles with lower error.}
	\label{fig:trainperformance}
\end{figure*}

We perform several analyses to evaluate our model given the limited training data. Figure \@ \ref{fig:trainperformance}a shows that there is no obvious correlation between the MSE and the number of basis atoms within unit cells among training, validation, and test datasets (additional statistics in are available in the \hyperref[sec:SIE(3)NNdetail]{Supplementary Material}). The overall test set error is higher compared to the training set but similar to that of the validation set, suggesting good generalizability. We also present the MSE as a function of different elements (Figure \ref{fig:trainperformance}b) and observe comparable error levels, indicating balanced prediction. Lastly, we compute the average phonon frequency $\bar{\omega} = \tfrac{\int \mathrm{d}\omega \, g(\omega) \, \omega }{\int \mathrm{d}\omega \, g(\omega)}$ for both E(3)NN-predicted and DFPT ground-truth spectra, which show excellent agreement on the test set (Figure \ref{fig:trainperformance}c); specifically, for 70\% of the testing samples, the relative error is below 10\%. This strongly suggests the capability of our model to predict phonon DoS.

To visualize the model performance, we plot 7 randomly selected examples from the test set in each error quartile in Figure \ref{fig:trainperformance}d, with rows 1 through 4 corresponding to the 1$^{\text{st}}$ quartile with highest agreement through the 4$^{\text{th}}$ quartile with lowest agreement, respectively. Additional examples are plotted in the \hyperref[sec:morepredictedphonondos]{Supplementary Material}. The predicted DoS in the 1$^{\text{st}}$ and 2$^{\text{nd}}$ quartiles show excellent agreement with DFPT calculations by reproducing fine features, while the 3$^{\text{rd}}$ and 4$^{\text{th}}$ quantiles show good or acceptable agreement by capturing main features. For instance, the predicted DoS of NaPF\textsubscript{6}, CdCN\textsubscript{2}, and NaCNO capture the energy of acoustic and optical phonon branches well but mispredict the relative amplitudes of certain peaks. Nonetheless, the phonon bandgap, a key quantity to determine phonon-phonon scattering, can be accurately extracted. Similarly, for KSbO\textsubscript{3} and  Li\textsubscript{2}GeO\textsubscript{3}, the predictions exhibit broadband DoS distribution, agreeing with DFPT calculations. A large discrepancy can be seen for RuS\textsubscript{2}, yet the bandwidth agreement is still good although the 0.099 MSE is among the largest errors in the test set (Figure \ref{fig:trainperformance}a). The good test set performance and generalizability suggest the suitability of our model to predict phonon DoS for a broad range of new materials. 


{\color{revision}
\section{Comparison to Experimental Data}
\label{sec:SICompExp}

We compare the E(3)NN predictions in 6 materials with experimental DoS data available from inelastic scattering (Figure \ref{fig:SIexpcomp}). Given the disorder and anharmonic effects in a measured sample, disagreement between DFPT calculations and measured data can happen. As a result, lower agreement is expected between experimental and E(3)NN-predicted DoS since the ground-truths are based on DFPT calculations. 

Although the E(3)NN-predicted DoS do not match the fine features of the experimental spectra, several key features (peak positions, gaps, and energy bandwidths) are still well-predicted and can be valuable in guiding experimental planning, which can serve as useful guidance for planning inelastic neutron and x-ray scattering measurements, where experimental resources are largely limited to national laboratory facilities.

\begin{figure}[t]
	\centering
	\includegraphics[width=\linewidth]{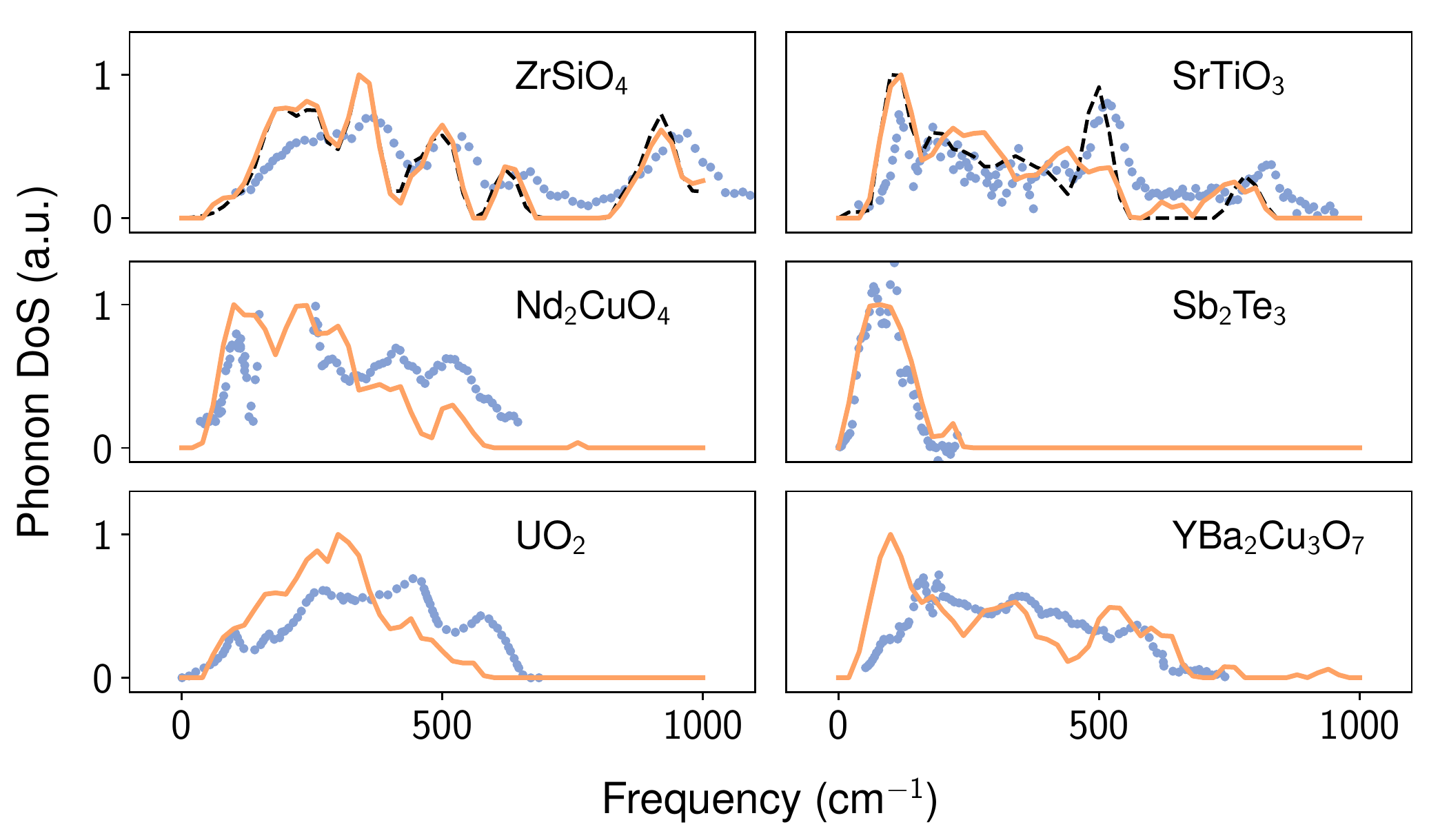}
	\caption{Comparison between E(3)NN model predictions (orange curves) and inelastic scattering DoS data (blue dots), reproduced from literature \cite{choudhury2008large,lynn1991phonon,sumarlin1993phonon,rauh1981generalized,pang2014phonon,chaplot2006phonon,nipko1997inelastic}.  ZrSiO\textsubscript{4} was in the training set and SrTiO\textsubscript{3} in the test set (black dashed lines denote corresponding DFPT results \cite{petretto2018high}). The remaining examples were absent in all datasets used for training, validation, and testing, and contain two unseen elements, Nd and U.}
	\label{fig:SIexpcomp}
\end{figure}
}

{\color{revision}

\section{Predicting phonon properties in alloy systems}

One of the most important applications of our prediction model may lie in the predictive power in alloy systems, particularly crystalline alloys with substitutional disorders. For example, given a binary alloy with composition $\mathrm{A}_{p}\mathrm{B}_{1-p} (0\leq p \leq 1)$, the input alloy encoding vector can take the following form 
\begin{equation}
\begin{split}
x_{\text{alloy}} = [0,\ldots,p m_{\text{A}},\ldots,(1-p) m_\text{{B}},\ldots,0],
\end{split}
\label{eq:alloy}
\end{equation}
where the two-hot encoding $pm_{\text{A}}$ and $(1-p)m_\text{{B}}$ are located at the vector indices corresponding the atomic numbers of A and B, respectively, weighted by composition. With this definition of Eq.\@ \ref{eq:alloy}, it can be directly reduced to pure phase one-hot encoding A (or B) by simply setting $p=1$ (or $p=0$), and it can be generalized to more complicated alloys directly. In fact, Eq.\@ \ref{eq:alloy} contains the essence of VCA with both mass average $m_{\text{VCA}}=pm_\text{A}+(1-p)m_\text{B}$ and scattering potential average $V_{\text{VCA}}=pV_\text{A}+(1-p)V_\text{B}$, but combined into one equation: the mass effect is contained in the numerical values of $pm_{\text{A}}$ and $(1-p)m_\text{{B}}$, and the potential effect is encoded by turning on the vector indices that correspond to the atomic species A and B. Since the computation of pure phase and alloy differs only by atomic embedding method, the alloy calculation does not generate any additional computational cost. 

We demonstrate the power of this approach with the alloy $\mathrm{Mg}_{3}\mathrm{Sb}_{2(1-p)}\mathrm{Bi}_{2p}$ with $p\in[0,1]$. The model evaluation is done by the aforementioned two-hot encoding for Sb on top of the structure Mg\textsubscript{3}Bi\textsubscript{2}, with simultaneously interpolating the lattice constants in-between values of two limit structures (namely, Mg\textsubscript{3}Bi\textsubscript{2} for $p=1$ and Mg\textsubscript{3}Sb\textsubscript{2} for $p=0$) according to the composition. In this case, both input vectors and the structure are recovered to the pure phase Mg\textsubscript{3}Sb\textsubscript{2} when $p=0$ and vice versa. We compute the phonon DoS in alloy Mg\textsubscript{3}Sb\textsubscript{0.5}Bi\textsubscript{1.5} and compare it with VCA calculations, as shown in Figure \ref{fig:alloy}, where the peak positions and magnitudes are both well predicted. The E(3)NN model used to evaluate this alloy system is trained with an additional Mg\textsubscript{3}Bi\textsubscript{2} phonon DoS curated from \cite{agne2018heat} compared with the model used in the rest of this paper, \textcolor{revision2}{while Mg\textsubscript{3}Sb\textsubscript{2} has already been included in the original training set}.

\begin{figure}[t]
	\centering
	\includegraphics[width=\linewidth]{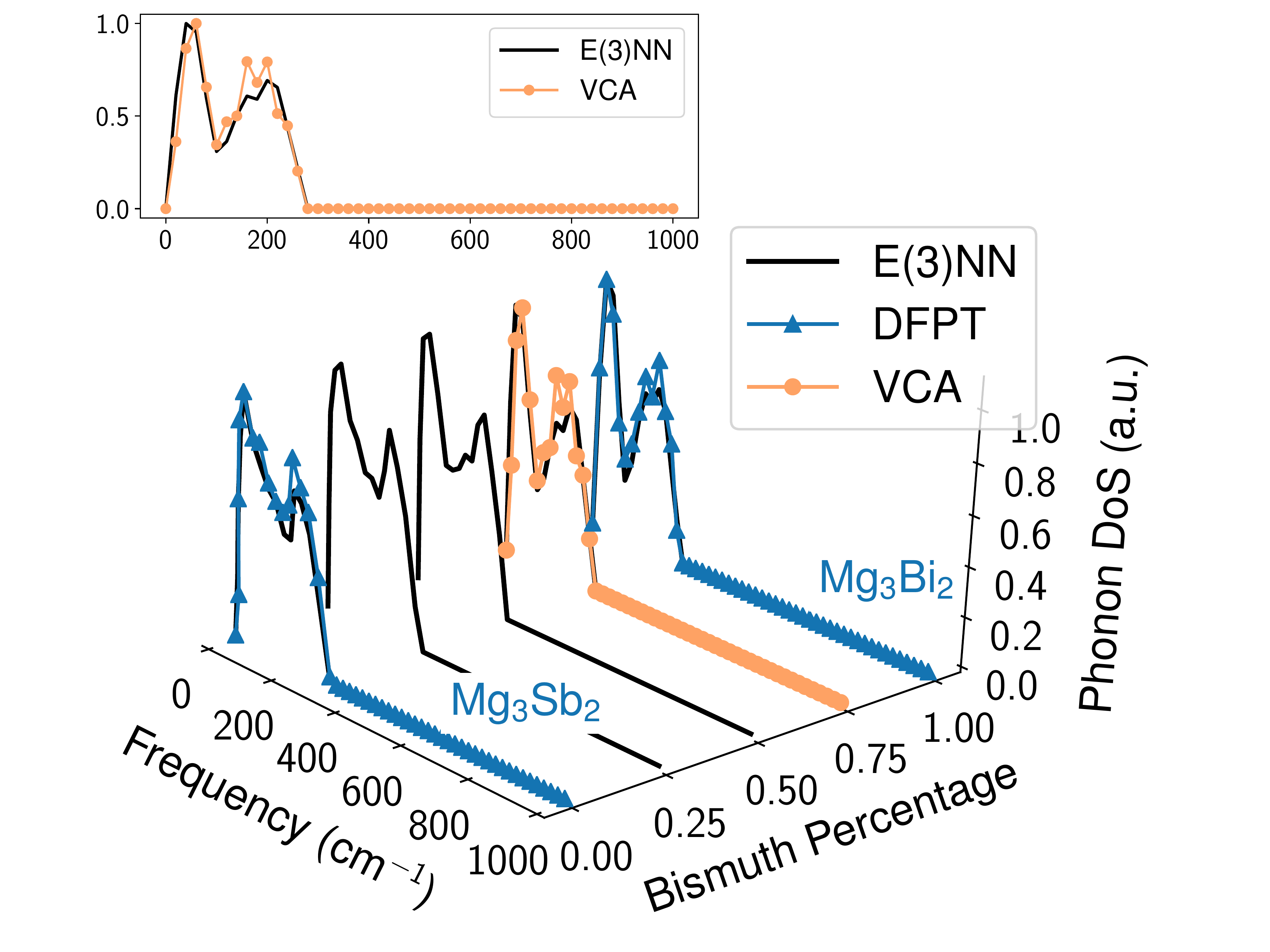}
	\caption{Comparison between E(3)NN model predictions and virtual crystal approximation (VCA) calculations. E(3)NN model can predict phonon DoS for the alloy $\mathrm{Mg}_{3}\mathrm{Sb}_{2(1-p)}\mathrm{Bi}_{2p}$ of continuous $p$ with two-hot weighted encoding. The triangle-marked curves indicate DFPT results for Mg\textsubscript{3}Sb\textsubscript{2} and Mg\textsubscript{3}Bi\textsubscript{2} curated from \cite{petretto2018high} and \cite{agne2018heat}, respectively. The circle-marked curve represent VCA calculations for Mg\textsubscript{3}Sb\textsubscript{0.5}Bi\textsubscript{1.5} ($p=0.75$). The inset figure shows the front view for E(3)NN and VCA comparison.}
	\label{fig:alloy}
\end{figure}

}

\section{High-throughput phonon DoS and specific heat capacity predictions}

\begin{figure*}
	\centering
	\includegraphics[width=\linewidth]{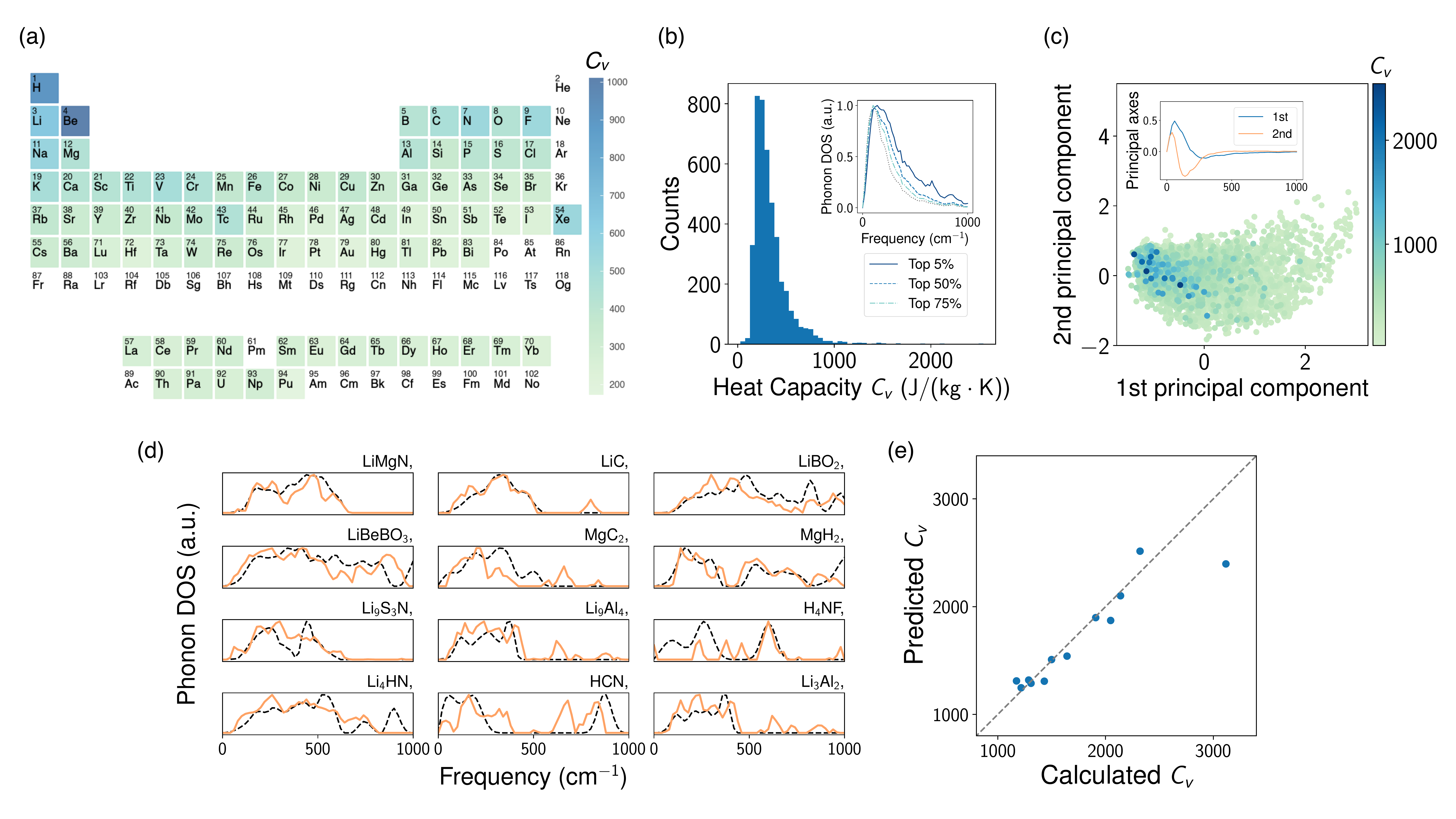}
	\caption{The search for high specific heat capacity ($C_{V}$) materials. (a) Periodic table colored by average $C_{V}$ of materials containing each element. (b) Histogram showing distribution of $C_{V}$ evaluated from E(3)NN-predicted phonon DoS. The inset illustrates average phonon DoS of materials with highest $C_{V}$. (c) Distribution of predicted phonon DoS along the first two principal components, colored by the E(3)NN-predicted $C_{V}$ magnitudes. The inset shows first two principal axes in the original frequency basis. (d) Comparison between E(3)NN-predicted and DFPT-computed phonon DoS, the dashed black curves represent DFPT results. (e) Two-dimensional histogram comparing specific heat capacities evaluated from E(3)NN-predicted and DFPT-calculated phonon DoS.}
	\label{fig:heatcapmatdiscover}
\end{figure*}

We apply the predictive model on 4,346 unseen crystal structures without ground-truth DoS from the Materials Project \cite{jain2013commentary}. The data consistency check is performed (\hyperref[sec:SIheatcapmetrics]{Supplementary Material}), showing a reasonable trend of an elastic spring model. Figure \ref{fig:heatcapmatdiscover}a illustrates the average phononic specific heat capacity $C_V$ of crystalline solids containing a given element, using the relation \cite{lee1995ab}
\begin{equation}
\begin{split}
    C_V(T)=\tfrac{k_{B}}{m_{tot}}\int_{0}^{\infty} \left(\tfrac{\hbar\omega}{2k_{B}T}\right)^{2}\mathrm{csch}^{2}\left(\tfrac{\hbar\omega}{2k_{B}T}\right)g(\omega)d\omega,
\end{split}
\label{eq:heatcapacity}
\end{equation}
where $m_{tot}$ is the total mass of all $N$ atoms in the unit cell, and the phonon DoS is normalized such that $\int g(\omega)d\omega = 3N$. Materials containing light elements tend to have high heat capacity, which is reasonable. The distribution of $C_V$ evaluated from Eq.\@ \ref{eq:heatcapacity} is shown in Figure \ref{fig:heatcapmatdiscover}b, where $\sim 2\%$ of materials show a $C_V$ greater than 1,000 $\mathrm{J/(kg \cdot K)}$. The inset shows the average phonon DoS of highest-$C_V$ materials. Materials with higher $C_V$ appear to have high spectral weight at higher energies, consistent with expectation. This trend is also noticed by inspecting the scatter plot of phonon DoS along the first two principal components (Figure \ref{fig:heatcapmatdiscover}c), where high heat capacity materials appear clustered with respect to the first principal axis. The first principal axis has a broad negative peak extending to high energies; thus, the clustering of high $C_{V}$ materials in the negative first principal direction parallels the shift of their phonon DoS towards higher energies.

To validate our model's predictions of high $C_V$ materials, we select 12 materials with ultrahigh predicted $C_V$ and carry out independent DFPT calculations. Since the maximum frequency of the training data was set to 1,000 cm$^{-1}$, which is sufficient for the majority of materials, the $C_V$ evaluated by DFPT was also cut off at 1,000 cm$^{-1}$ for fair comparison. The DoS  comparisons between E(3)NN and DFPT in these high-$C_{V}$ materials are shown in \ref{fig:heatcapmatdiscover}d, where satisfactory agreements are achieved in most examples except for H\textsubscript{4}NF and HCN. The $C_V$ at room temperature $T=293.15\,\mathrm{K}$ evaluated from E(3)NN predictions and DFPT calculations are plotted in Figure \ref{fig:heatcapmatdiscover}e, showing excellent agreement for most materials. Values of the plotted $C_{V}$ are presented in Table \ref{tab:SIcompareheatcap} in the \hyperref[sec:fullrangedos]{Supplementary Material}, together with $C_{V}$ computed from full energy ranges for comparison. We attribute the large discrepancy in hydrogen- and lithium-rich materials to the electrostatic effect of hydrogen and lithium bondings beyond mass effect \cite{shahi2014hydrogen}, while the current model mainly considers the mass effect. 

\section{Discussion}

In this work, we present a machine learning-based predictive model to directly acquire the high-dimensional material property of phonon DoS, using only ``first-principles'' inputs, namely atomic species and positions. Due to their equivariance, Euclidean neural networks are able to capture the symmetries of the input crystal, making them data-efficient. A small training set of only 1,200 examples is sufficient to generate meaningful predictions, outperforming a well-trained convolutional neural network even with data augmentation (\hyperref[tab:SIconvneuralnet]{Supplementary Material}). \textcolor{revision}{One limitation of the current model is identified as samples with elastic strain, potentially related to the fact that we confined output energy range to positive values (\hyperref[fig:SI_STO_strain]{Supplementary Material})}. Even so, Euclidean neural networks can be applied to predicting broader properties in crystalline solids, where there are often issues of data scarcity. \textcolor{revision}{Most importantly, our atomic embedding approach offers an extremely efficient way to compute phonon DoS in alloys, where the alloy prediction has the same computational cost as pure phase by simply changing the 1-hot embedding to a weighted multi-hot embedding.}

In contrast to \textit{ab initio} calculations and inelastic scattering that acquire phonon DoS deterministically, a ML model is data-driven and probabilistic in nature. It is thus impractical to fully rely on a ML-based predictive model to acquire materials properties without further validation. However, the power of the ML approach goes far beyond obtaining property magnitudes at the individual material level. From a materials design perspective, ML demonstrates extremely high efficiency in rapidly screening candidates with a target property. In our case, the prediction of phonon DoS on the 4,346 unseen materials can be done in less than 30 minutes on a single entry-level GPU. From a property optimization perspective, instead of measuring an individual material with high precision, the ML approach searches outputs high-performance candidates in a batch, offering more choices. From an experimental perspective, the ML model can be valuable in guiding the experimental planning with limited national facility resources. From an application point-of-view, the Dulong-Petit law poses a grand challenge in searching for promising materials for thermal storage \cite{henry2020five}, and a highly efficient approach can further support an inverse design from property to desirable structures. In sum, our model provides a promising framework to enable high-throughput screening and guide experimental planning for materials with exceptional thermal properties. It further sheds light on elucidating the fundamental links between symmetry, structure, and elementary excitations in condensed matter. 

\section*{Acknowledgments}
Z.C., N.A. and M.L. thank C.H. Rycroft and J.W. Lynn for helpful discussions.  Z.C., N.A., and M.L.\@ acknowledge the support from U.S.\@ DOE, Office of Science (SC), Basic Energy Sciences (BES), award No.\@ DE-SC0020148. N.A.\@ acknowledges the support of the National Science Foundation Graduate Research Fellowship Program under Grant No. 1122374. T.E.S. acknowledges support from the Laboratory Directed Research and Development Program of Lawrence Berkeley National Laboratory under U.S. Department of Energy Contract No. DE-AC02-05CH11231. Y.T. thanks K. Van Vliet for computational support.

\bibliography{2020_phonon_dos}

\begin{thebibliography}{10}
\providecommand{\url}[1]{\texttt{#1}}
\providecommand{\urlprefix}{URL }

\bibitem{ramprasad2017machine}
R.~Ramprasad, R.~Batra, G.~Pilania, A.~Mannodi-Kanakkithodi, C.~Kim,
\newblock \emph{npj Computational Materials} \textbf{2017}, \emph{3}, 1 1.

\bibitem{schmidt2019recent}
J.~Schmidt, M.~R. Marques, S.~Botti, M.~A. Marques,
\newblock \emph{npj Computational Materials} \textbf{2019}, \emph{5}, 1 1.

\bibitem{raccuglia2016machine}
P.~Raccuglia, K.~C. Elbert, P.~D. Adler, C.~Falk, M.~B. Wenny, A.~Mollo,
  M.~Zeller, S.~A. Friedler, J.~Schrier, A.~J. Norquist,
\newblock \emph{Nature} \textbf{2016}, \emph{533}, 7601 73.

\bibitem{oliynyk2016high}
A.~O. Oliynyk, E.~Antono, T.~D. Sparks, L.~Ghadbeigi, M.~W. Gaultois,
  B.~Meredig, A.~Mar,
\newblock \emph{Chemistry of Materials} \textbf{2016}, \emph{28}, 20 7324.

\bibitem{gomez2018automatic}
R.~G{\'o}mez-Bombarelli, J.~N. Wei, D.~Duvenaud, J.~M. Hern{\'a}ndez-Lobato,
  B.~S{\'a}nchez-Lengeling, D.~Sheberla, J.~Aguilera-Iparraguirre, T.~D.
  Hirzel, R.~P. Adams, A.~Aspuru-Guzik,
\newblock \emph{ACS central science} \textbf{2018}, \emph{4}, 2 268.

\bibitem{LIU2017discovery_design}
Y.~Liu, T.~Zhao, W.~Ju, S.~Shi,
\newblock \emph{Journal of Materiomics} \textbf{2017}, \emph{3}, 3 159 ,
  high-throughput Experimental and Modeling Research toward Advanced Batteries.

\bibitem{Alan2018Chimera}
F.~Häse, L.~M. Roch, A.~Aspuru-Guzik,
\newblock \emph{Chem. Sci.} \textbf{2018}, \emph{9} 7642.

\bibitem{Granda2018}
J.~M. Granda, L.~Donina, V.~Dragone, D.-L. Long, L.~Cronin,
\newblock \emph{Nature} \textbf{2018}, \emph{559}, 7714 377.

\bibitem{pilania2013accelerating}
G.~Pilania, C.~Wang, X.~Jiang, S.~Rajasekaran, R.~Ramprasad,
\newblock \emph{Scientific reports} \textbf{2013}, \emph{3}, 1 1.

\bibitem{isayev2017universal}
O.~Isayev, C.~Oses, C.~Toher, E.~Gossett, S.~Curtarolo, A.~Tropsha,
\newblock \emph{Nature communications} \textbf{2017}, \emph{8}, 1 1.

\bibitem{xie2018crystal}
T.~Xie, J.~C. Grossman,
\newblock \emph{Physical review letters} \textbf{2018}, \emph{120}, 14 145301.

\bibitem{chen2019graph}
C.~Chen, W.~Ye, Y.~Zuo, C.~Zheng, S.~P. Ong,
\newblock \emph{Chemistry of Materials} \textbf{2019}, \emph{31}, 9 3564.

\bibitem{carrete2014finding}
J.~Carrete, W.~Li, N.~Mingo, S.~Wang, S.~Curtarolo,
\newblock \emph{Physical Review X} \textbf{2014}, \emph{4}, 1 011019.

\bibitem{van2016high}
A.~van Roekeghem, J.~Carrete, C.~Oses, S.~Curtarolo, N.~Mingo,
\newblock \emph{Physical Review X} \textbf{2016}, \emph{6}, 4 041061.

\bibitem{tawfik2020predicting}
S.~A. Tawfik, O.~Isayev, M.~J. Spencer, D.~A. Winkler,
\newblock \emph{Advanced Theory and Simulations} \textbf{2020}, \emph{3}, 2
  1900208.

\bibitem{mortazavi2020machine}
B.~Mortazavi, E.~V. Podryabinkin, S.~Roche, T.~Rabczuk, X.~Zhuang, A.~V.
  Shapeev,
\newblock \emph{Materials Horizons} \textbf{2020}.

\bibitem{ward2016general}
L.~Ward, A.~Agrawal, A.~Choudhary, C.~Wolverton,
\newblock \emph{npj Computational Materials} \textbf{2016}, \emph{2}, 1 1.

\bibitem{zhuo2018predicting}
Y.~Zhuo, A.~Mansouri~Tehrani, J.~Brgoch,
\newblock \emph{The journal of physical chemistry letters} \textbf{2018},
  \emph{9}, 7 1668.

\bibitem{dong2019bandgap}
Y.~Dong, C.~Wu, C.~Zhang, Y.~Liu, J.~Cheng, J.~Lin,
\newblock \emph{npj Computational Materials} \textbf{2019}, \emph{5}, 1 1.

\bibitem{meredig2018can}
B.~Meredig, E.~Antono, C.~Church, M.~Hutchinson, J.~Ling, S.~Paradiso,
  B.~Blaiszik, I.~Foster, B.~Gibbons, J.~Hattrick-Simpers, et~al.,
\newblock \emph{Molecular Systems Design \& Engineering} \textbf{2018},
  \emph{3}, 5 819.

\bibitem{stanev2018machine}
V.~Stanev, C.~Oses, A.~G. Kusne, E.~Rodriguez, J.~Paglione, S.~Curtarolo,
  I.~Takeuchi,
\newblock \emph{npj Computational Materials} \textbf{2018}, \emph{4}, 1 1.

\bibitem{andrejevic2020machine}
N.~Andrejevic, J.~Andrejevic, C.~H. Rycroft, M.~Li,
\newblock \emph{arXiv preprint arXiv:2003.00994} \textbf{2020}.

\bibitem{claussen2020detection}
N.~Claussen, B.~A. Bernevig, N.~Regnault,
\newblock \emph{Physical Review B} \textbf{2020}, \emph{101}, 24 245117.

\bibitem{scheurer2020unsupervised}
M.~S. Scheurer, R.-J. Slager,
\newblock \emph{Physical Review Letters} \textbf{2020}, \emph{124}, 22 226401.

\bibitem{li2015molecular}
Z.~Li, J.~R. Kermode, A.~De~Vita,
\newblock \emph{Physical review letters} \textbf{2015}, \emph{114}, 9 096405.

\bibitem{chmiela2017machine}
S.~Chmiela, A.~Tkatchenko, H.~E. Sauceda, I.~Poltavsky, K.~T. Sch{\"u}tt, K.-R.
  M{\"u}ller,
\newblock \emph{Science advances} \textbf{2017}, \emph{3}, 5 e1603015.

\bibitem{kruglov2017energy}
I.~Kruglov, O.~Sergeev, A.~Yanilkin, A.~R. Oganov,
\newblock \emph{Scientific reports} \textbf{2017}, \emph{7}, 1 1.

\bibitem{glielmo2017accurate}
A.~Glielmo, P.~Sollich, A.~De~Vita,
\newblock \emph{Physical Review B} \textbf{2017}, \emph{95}, 21 214302.

\bibitem{botu2017machine}
V.~Botu, R.~Batra, J.~Chapman, R.~Ramprasad,
\newblock \emph{The Journal of Physical Chemistry C} \textbf{2017}, \emph{121},
  1 511.

\bibitem{legrain2018vibrational}
F.~Legrain, A.~van Roekeghem, S.~Curtarolo, J.~Carrete, G.~K. Madsen, N.~Mingo,
\newblock \emph{Journal of Chemical Information and Modeling} \textbf{2018},
  \emph{58}, 12 2460.

\bibitem{zhang2019active}
L.~Zhang, D.-Y. Lin, H.~Wang, R.~Car, E.~Weinan,
\newblock \emph{Physical Review Materials} \textbf{2019}, \emph{3}, 2 023804.

\bibitem{Wu2019interfacial}
Y.-J. Wu, L.~Fang, Y.~Xu,
\newblock \emph{npj Computational Materials} \textbf{2019}, \emph{5}, 1 56.

\bibitem{Giustino2017elph}
F.~Giustino,
\newblock \emph{Rev. Mod. Phys.} \textbf{2017}, \emph{89} 015003.

\bibitem{bardeen1957theory}
J.~Bardeen, L.~N. Cooper, J.~R. Schrieffer,
\newblock \emph{Physical review} \textbf{1957}, \emph{108}, 5 1175.

\bibitem{baroni2001phonons}
S.~Baroni, S.~De~Gironcoli, A.~Dal~Corso, P.~Giannozzi,
\newblock \emph{Reviews of Modern Physics} \textbf{2001}, \emph{73}, 2 515.

\bibitem{Seyf2017Ase}
H.~R. Seyf, L.~Yates, T.~L. Bougher, S.~Graham, B.~A. Cola, T.~Detchprohm,
  M.-H. Ji, J.~Kim, R.~Dupuis, W.~Lv, A.~Henry,
\newblock \emph{npj Computational Materials} \textbf{2017}, \emph{3}, 1 49.

\bibitem{thomas2018tensor}
N.~{Thomas}, T.~{Smidt}, S.~{Kearnes}, L.~{Yang}, L.~{Li}, K.~{Kohlhoff},
  P.~{Riley},
\newblock \emph{arXiv e-prints} \textbf{2018}, arXiv:1802.08219.

\bibitem{kondor2018clebsch}
R.~Kondor, Z.~Lin, S.~Trivedi,
\newblock In \emph{Advances in Neural Information Processing Systems 32}.
  \textbf{2018} 10117--10126,
\newblock \urlprefix\url{https://dl.acm.org/doi/proceedings/10.5555/3326943}.

\bibitem{3dsteerable}
M.~Weiler, M.~Geiger, M.~Welling, W.~Boomsma, T.~Cohen,
\newblock In \emph{Advances in Neural Information Processing Systems 32}.
  \textbf{2018} 10402--10413,
\newblock \urlprefix\url{https://dl.acm.org/doi/proceedings/10.5555/3326943}.

\bibitem{e3nn_2020_3724963}
M.~Geiger, T.~Smidt, B.~K. Miller, W.~Boomsma, K.~Lapchevskyi, M.~Weiler,
  M.~Tyszkiewicz, B.~Dice, J.~Frellsen, S.~Sanborn, M.~Alby,
\newblock \texttt{e3nn}: a modular framework for euclidean neural networks,
  github.com/e3nn/e3nn.

\bibitem{Tess2020Euclidean}
T.~E. {Smidt}, M.~{Geiger}, B.~K. {Miller},
\newblock \emph{arXiv e-prints} \textbf{2020}, arXiv:2007.02005.

\bibitem{petretto2018high}
G.~Petretto, S.~Dwaraknath, H.~P. Miranda, D.~Winston, M.~Giantomassi, M.~J.
  Van~Setten, X.~Gonze, K.~A. Persson, G.~Hautier, G.-M. Rignanese,
\newblock \emph{Scientific data} \textbf{2018}, \emph{5} 180065.

\bibitem{e3nnQM9}
B.~K. {Miller}, M.~{Geiger}, T.~E. {Smidt}, F.~{No{\'e}},
\newblock \emph{arXiv e-prints} \textbf{2020}, arXiv:2008.08461.

\bibitem{choudhury2008large}
N.~Choudhury, E.~J. Walter, A.~I. Kolesnikov, C.-K. Loong,
\newblock \emph{Physical Review B} \textbf{2008}, \emph{77}, 13 134111.

\bibitem{lynn1991phonon}
J.~Lynn, I.~Sumarlin, D.~Neumann, J.~Rush, J.~Peng, Z.~Li,
\newblock \emph{Physical review letters} \textbf{1991}, \emph{66}, 7 919.

\bibitem{sumarlin1993phonon}
I.~Sumarlin, J.~Lynn, D.~Neumann, J.~Rush, C.~Loong, J.~Peng, Z.~Li,
\newblock \emph{Physical Review B} \textbf{1993}, \emph{48}, 1 473.

\bibitem{rauh1981generalized}
H.~Rauh, R.~Geick, H.~Kohler, N.~Nucker, N.~Lehner,
\newblock \emph{Journal of Physics C: Solid State Physics} \textbf{1981},
  \emph{14}, 20 2705.

\bibitem{pang2014phonon}
J.~W. Pang, A.~Chernatynskiy, B.~C. Larson, W.~J. Buyers, D.~L. Abernathy,
  K.~J. McClellan, S.~R. Phillpot,
\newblock \emph{Physical Review B} \textbf{2014}, \emph{89}, 11 115132.

\bibitem{chaplot2006phonon}
S.~Chaplot, L.~Pintschovius, N.~Choudhury, R.~Mittal,
\newblock \emph{Physical Review B} \textbf{2006}, \emph{73}, 9 094308.

\bibitem{nipko1997inelastic}
J.~Nipko, C.-K. Loong,
\newblock \emph{Physica B: Condensed Matter} \textbf{1997}, \emph{241} 415.

\bibitem{agne2018heat}
M.~T. Agne, K.~Imasato, S.~Anand, K.~Lee, S.~K. Bux, A.~Zevalkink, A.~J.
  Rettie, D.~Y. Chung, M.~G. Kanatzidis, G.~J. Snyder,
\newblock \emph{Materials Today Physics} \textbf{2018}, \emph{6} 83.

\bibitem{jain2013commentary}
A.~Jain, S.~P. Ong, G.~Hautier, W.~Chen, W.~D. Richards, S.~Dacek, S.~Cholia,
  D.~Gunter, D.~Skinner, G.~Ceder, et~al.,
\newblock \emph{Apl Materials} \textbf{2013}, \emph{1}, 1 011002.

\bibitem{lee1995ab}
C.~Lee, X.~Gonze,
\newblock \emph{Physical Review B} \textbf{1995}, \emph{51}, 13 8610.

\bibitem{shahi2014hydrogen}
A.~Shahi, E.~Arunan,
\newblock \emph{Physical Chemistry Chemical Physics} \textbf{2014}, \emph{16},
  42 22935.

\bibitem{henry2020five}
A.~Henry, R.~Prasher, A.~Majumdar,
\newblock \emph{Nature Energy} \textbf{2020}, 1--3.

\bibitem{Ong2013}
S.~P. Ong, W.~D. Richards, A.~Jain, G.~Hautier, M.~Kocher, S.~Cholia,
  D.~Gunter, V.~L. Chevrier, K.~A. Persson, G.~Ceder,
\newblock \emph{Computational Materials Science} \textbf{2013}, \emph{68} 314.

\bibitem{lynn1973lattice}
J.~Lynn, H.~Smith, R.~Nicklow,
\newblock \emph{Physical Review B} \textbf{1973}, \emph{8}, 8 3493.

\bibitem{hoffmann2019data}
J.~Hoffmann, L.~Maestrati, Y.~Sawada, J.~Tang, J.~M. Sellier, Y.~Bengio,
\newblock \emph{arXiv preprint arXiv:1909.00949} \textbf{2019}.

\end{thebibliography}
\bibliographystyle{MSP}

\onecolumngrid
\newpage


\setcounter{figure}{0}
\setcounter{equation}{0}
\setcounter{table}{0}
\renewcommand{\thetable}{S\arabic{table}}
\renewcommand{\thefigure}{S\arabic{figure}}
\renewcommand{\theequation}{S\arabic{equation}}


\begin{center}
    \large{\textbf{\textsc{Supplementary Material}}}
\end{center}

\section{General validity of predictions on unseen materials}
\label{sec:SIheatcapmetrics}

\begin{figure}[h]
	\centering
	\includegraphics[width=0.5\linewidth]{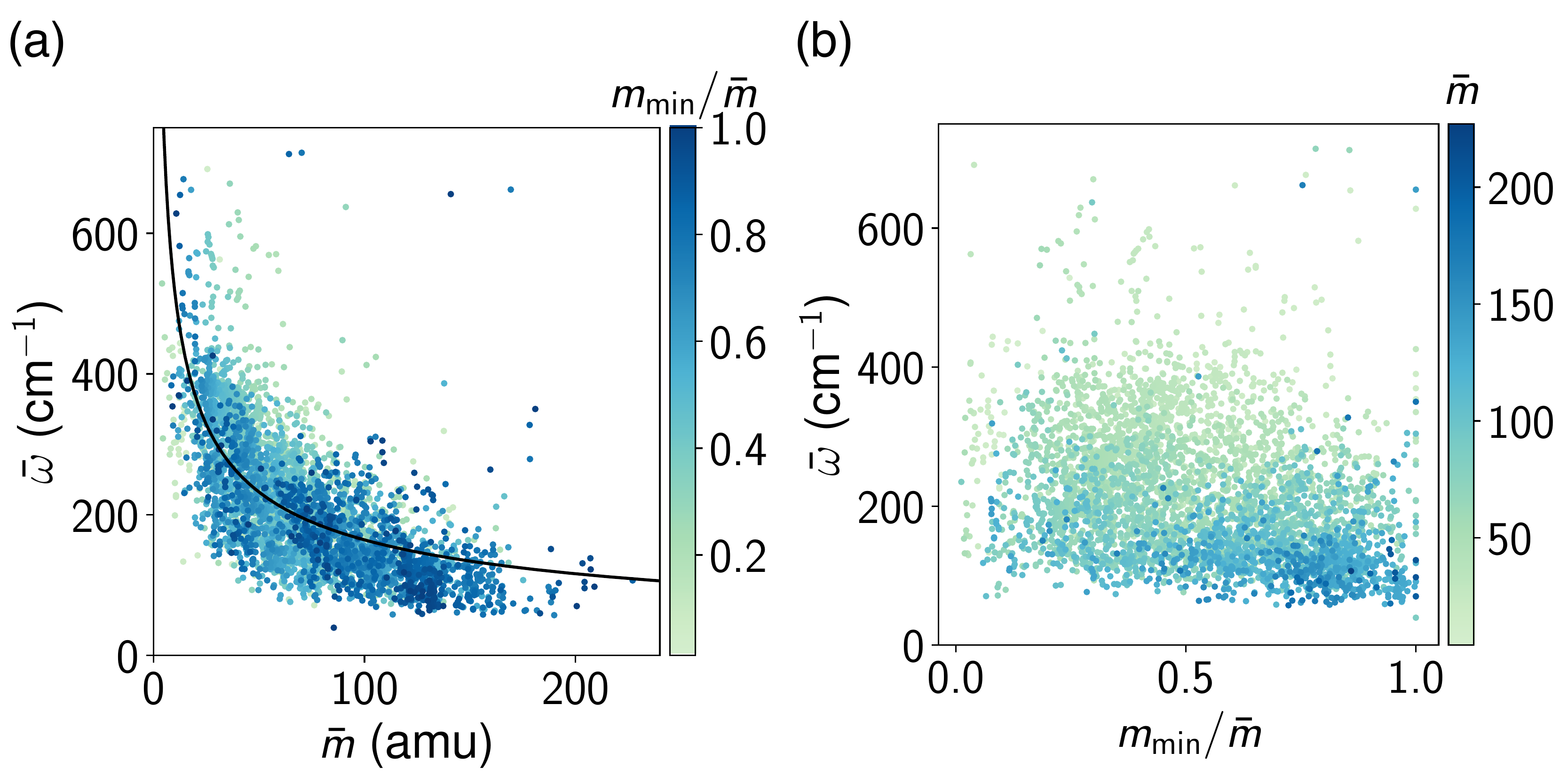}
	\caption{Evaluation of model predictions on unseen crystal structures. (a) Average frequency $\bar{\omega}$ versus average atomic mass $\bar{m}$. The black solid line represents the least squared fit to the hyperbolic relation $\bar{\omega}\sim\bar{m}^{-1/2}$. (b) The average frequency versus the ratio $m_{\min}/\bar{m}$ characterizing atomic mass non-uniformity.}
	\label{fig:SIheatcapmetrics}
\end{figure}

We apply the predictive model on 4,346 unseen crystal structures from the Materials Project \cite{jain2013commentary} with atomic site number $N\leq13$ in each unit cell (consistent with most materials presented in \cite{petretto2018high}). As a check, we plot the average phonon frequency $\bar{\omega}$ against the average atomic mass $\bar{m}=(\frac{1}{N}\sum_{i=1}^{N}\sqrt{m_{i}})^{2}$ in the unit cell (Figure \ref{fig:SIheatcapmetrics}a ). The predicted data fit well to a hyperbolic curve $\bar{\omega}=C\bar{m}^{-1/2}$, where the constant $C$ is a measure of the crystal rigidity. The reasonable distribution of rigidity supports the physical validity of our model for new materials. Moreover, we characterize the non-uniformity of atomic masses in each material by computing the ratio of the minimum mass $m_{\min}$ in a crystal to $\bar{m}$, where high $m_{\min}/\bar{m}$ tends to aggregate at lower $\bar{\omega}$ and higher $\bar{m}$ (Figure \ref{fig:SIheatcapmetrics}b), which agrees with the features in \cite{petretto2018high}. 

\section{Euclidean neural network in detail}
\label{sec:SIE(3)NNdetail}

As discussed in the main text, crystals are represented to the network as periodic graphs where relative distance vectors are edge attributes and atom features are stored at each node as shown in Figure \ref{fig:networkworkflow}. Crystal structures are converted into graphs using the \texttt{e3nn} \texttt{DataPeriodicNeighbors} class which uses the \texttt{pymatgen} \texttt{Structure} class to find atom neighbors while accounting for periodic boundary conditions \cite{Ong2013}. For each atom, all atoms around a given atom within a given cutoff radius (including the central atom) are considered ``neighbors'' and are used in the convolution operation. The features of each atom in the unit cell is stored in $x_{ai}$ where $a$ is the atom index and $i$ is a flattened representation index with features given by a representation list \texttt{Rs = [(N, 0, 1)]} which denotes \texttt{N} scalars. Representation lists in \texttt{e3nn} denote irreducible representations (irreps) of the group of 3D rotations and inversion, $O(3)$, and are lists of triples \texttt{(m, $L$, p)} denoting the number of copies or multiplicity \texttt{m} of features with rotation degree $L$ and parity \texttt{p} (1 for even parity, -1 for odd parity, and 0 for irreps of the group of just 3D rotations $SO(3)$). An irrep of rotation degree $L$ has $2 L + 1$ components. By flattening the representations into one index and using representation lists, we can efficient store and operate on geometric tensor quantities expressed in the irrep basis.

We employ the modified one-hot encoding for element types with atomic mass as magnitudes, namely $V_{vc}^{(0)}|_{c=Z_{v}}=m_{v}$ for site $v$ with atomic number $Z_{v}$ and mass $m_{v}$ (note the subscript $m\equiv1$ is omitted for order $l=0$). For example,
\begin{equation}
\begin{split}
    \text{H:}  \quad [1,0,\ldots,0,0],\qquad
    \text{He:} \quad [0,4,\ldots,0,0],\qquad
    \ldots ,   \qquad
    \text{Og:} \quad [0,0,\ldots,0,294].
\end{split}
\end{equation}

\begin{figure}[h]
    \centering
    \includegraphics[width=0.6\linewidth]{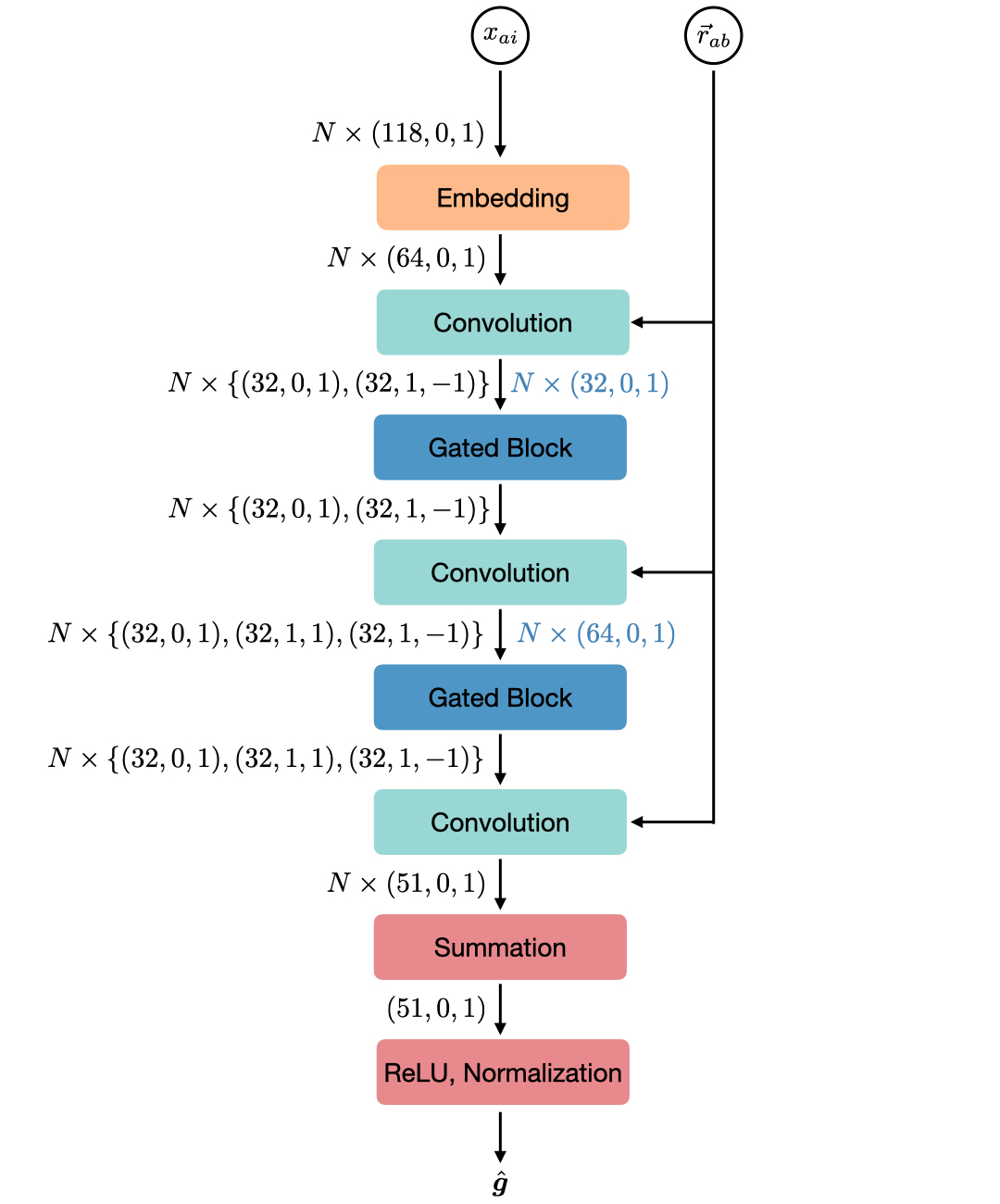}
    \caption{Architecture of the Euclidean neural network.}
    \label{fig:SInetworkarch}
\end{figure}

\noindent The encoded crystal graph is then passed into a E(3)NN as illustrated in Figure \ref{fig:SInetworkarch}. The architecture of the Euclidean neural networks used in this work is similar to that of a graph convolutional neural networks. To achieve Euclidean symmetry equivariance: 1) E(3)NN convolutional filters are functions of the radial distance vector between two points and composed of learned radial functions and spherical harmonics $W(\vec{r}) = R(|r|) Y_{lm}(\hat{r})$. 2) As a consequence of this filter choice, all inputs, intermediate data, and outputs are geometric tensors. 3) Therefore, all scalar operations (e.g. addition and multiplication) in the network must be replaced with general geometric tensor algebra. 4) Additionally, nonlinearities applied to geometric tensor data must also be replaced with equivariant equivalents. A feature that emerges from equivariance is that the symmetry of the outputs of Euclidean neural networks are guaranteed to have equal or higher symmetry than the inputs, which means that these networks are guaranteed to respect the space group symmetries (which are a subgroup of Euclidean symmetry) of input crystal geometry \cite{Tess2020Euclidean}.

To articulate network operations, we will use Einstein summation notation where repeated indices are implicitly summed over. A single layer of our network operates on input $x_{ai}$ and relative distance vectors of graph edges $\vec{r}_{ab}$,

\begin{align}
    x^{(q + 1)}_{ai} = \sigma( x^{(q)}_{bj} \otimes K_k(\vec{r}_{ab}))
\end{align}
where $\sigma$ is an equivariant nonlinearity and $\otimes$ signifies a tensor product where representation indices of inputs and filter are contracted using Clebsch-Gordan coefficients. We used a ``gated'' rotation equivariant nonlinearity, \texttt{GatedBlockParity} as implemented in \texttt{e3nn}, which was first introduced in Ref.~\cite{3dsteerable} and is extended in \texttt{e3nn} to handle parity (inversion). $K$ is the convolutional kernel which is composed of learned radial functions and spherical harmonics and Clebsch-Gordan coefficients are included in the kernel to yield the traditional channel out and in indices.
\begin{align}
    K_{ij}(\vec{r}_{ab}) = K_{abij} =  R_{w}(r_{ab}) Y_k(\hat{r}_{ab}) C_{ijk}
    \delta_{w, k \in\text{irrep}(w)}
\end{align}
where $\delta_{w, k \in \text{irrep}(w)}$ denotes that radial functions are shared for all components of a given irrep, e.g. the 5 components of a $L=2$ irrep share the same radial function, and $C_{ijk}$ are the Clebsch-Gordan coeffcients.

In tensor notation, a convolutional operation is written as
\begin{align}
    \text{Conv}(x_{bi},\vec{r}_{ab}) \coloneqq x_{bj} K_{ij}(\vec{r}_{ab}) = y_{ai}
\end{align}

We use convolutional filters up to $L \le 1$ and rotation order for intermediate features of $L \le 1$. 

The radial functions are dense (fully-connected) neural networks acting on a finite radial basis. For example, a two layer radial function would be expressed as
\begin{align}
    R(|r_{ab}|) = W_{kh} \sigma(W_{hq} B_{q}(|r_{ab}|))
\end{align}
where $B_q$ are the set of radial basis functions; in this work, we use a finite set of Gaussian radial basis functions. 

The first two convolutional layers generate $L=\{0,1\}$ atomic features and additional scalars to be used by following gated blocks for nonlinearizing $L=1$ pseudovectors \cite{e3nnQM9}. The final convolution operation yields atomic features of order $L=0$ on each atom. Finally, states at all sites within the unit cell are aggregated into a one-dimensional array $\sum_{a \in N} x_{ia}^{(q)}$. We then apply a ReLU activation and normalize (by dividing by the maximum intensity) to predict the phonon density of states (DoS), which is simply an array of 51 scalars. 

\section{Data pre-processing and preparation}

The phonon DoS dataset with 1,521 crystalline solids calculated from density functional perturbation theory (DFPT) by Petretto et al.\@ \cite{petretto2018high} is used for training and testing a predictive model in this study. We randomly split the entire dataset into training (80\%), validation (10\%), and test (10\%) sets. We manually curated 3 additional experimental phonon DoS from \cite{lynn1973lattice}, adding the Cu and Ag examples to the training set and Au to the test set, in order to provide the network examples of single-element compounds. The resulting training set had 1220 samples, and each of validation and test sets had 152 samples.  

The DFPT-calculated phonon DoS data has high energy resolution, requiring a large number of parameters in the neural network to fit the output dimension. Given limited training data, it is challenging to train a predictive model with too many trainable weights. To ensure a balanced output dimension and resolution while retaining the main features of the phonon DoS, we interpolated the smoothed spectrum in the energy range $0\le \omega\le 1000\, \mathrm{cm}^{-1}$ to 51 points. {\color{revision}The number 51 is well chosen in the sense is that it matches the scattering instrumental resolution. The sampled data points correspond to an energy segmentation of $\sim2.4\,\mathrm{meV}$ ($20\,\mathrm{cm}^{-1}$), which is roughly the same for typical inelastic scattering measurements with energy resolution $\sim2.0\,\mathrm{meV}$, below which the finer features are not distinguishable by any existing technique. For instance, the state-of-the-art inelastic scattering at NSLS-II has $\sim2.0\,\mathrm{meV}$ resolution. It should also be noticed that phonon DoS of some calculated materials in the dataset has negative frequencies components. According to \cite{petretto2018high}, those materials are not accompanied with thermodynamic properties and it is therefore found that around 1/6 of the total data has imaginary modes. However, since our model is confined to the energy range $0\le \omega\le 1000\, \mathrm{cm}^{-1}$ and is intrinsically unaware of the truncated information at negative frequencies. This could partially explain the poor behavior of our model in predicting strained samples as it is not trained to differentiate stable and unstable structures.}

\begin{figure}[h]
    \centering
    \includegraphics[width=0.35\linewidth]{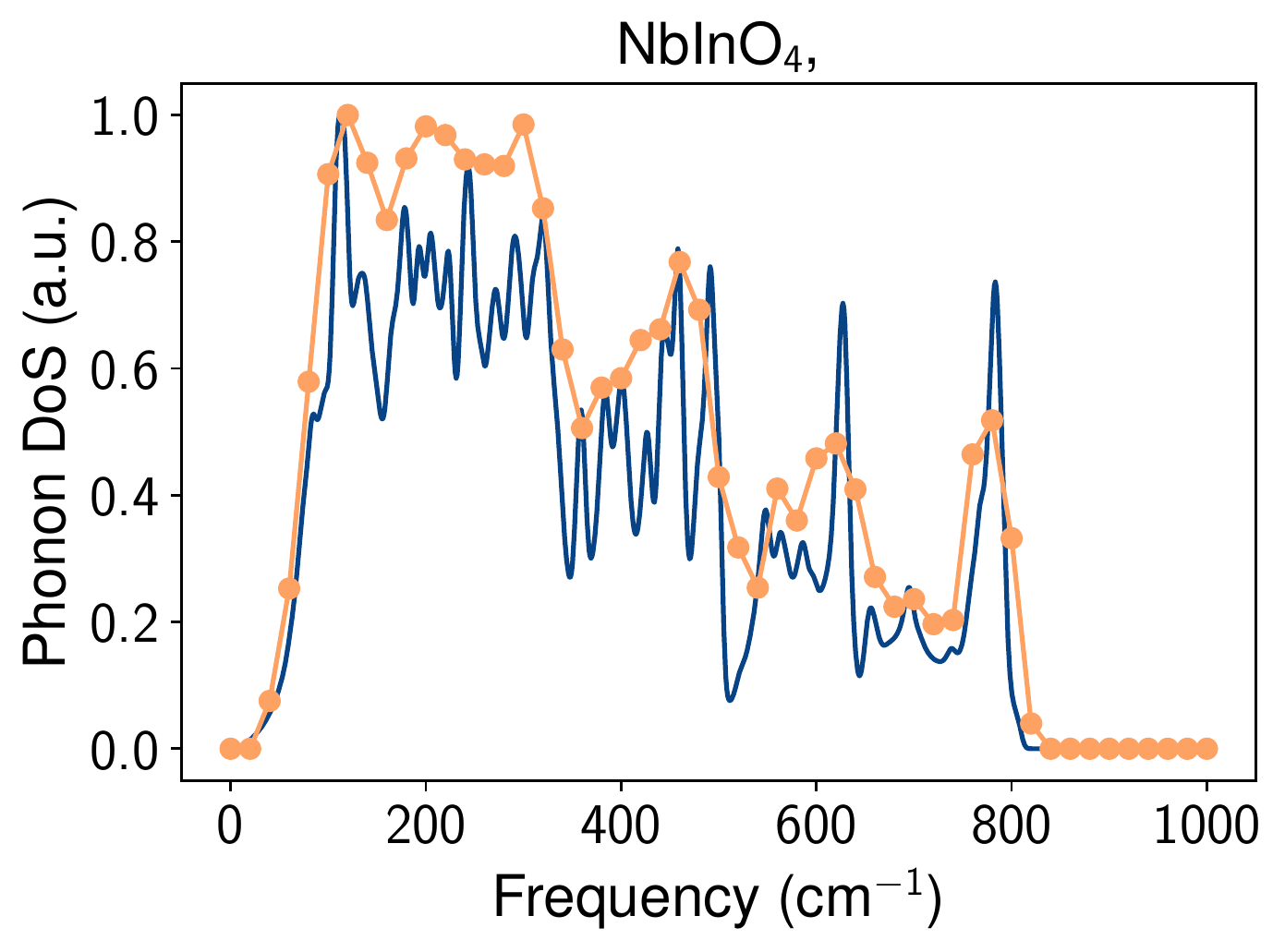}
    \hspace{1cm}
    \includegraphics[width=0.35\linewidth]{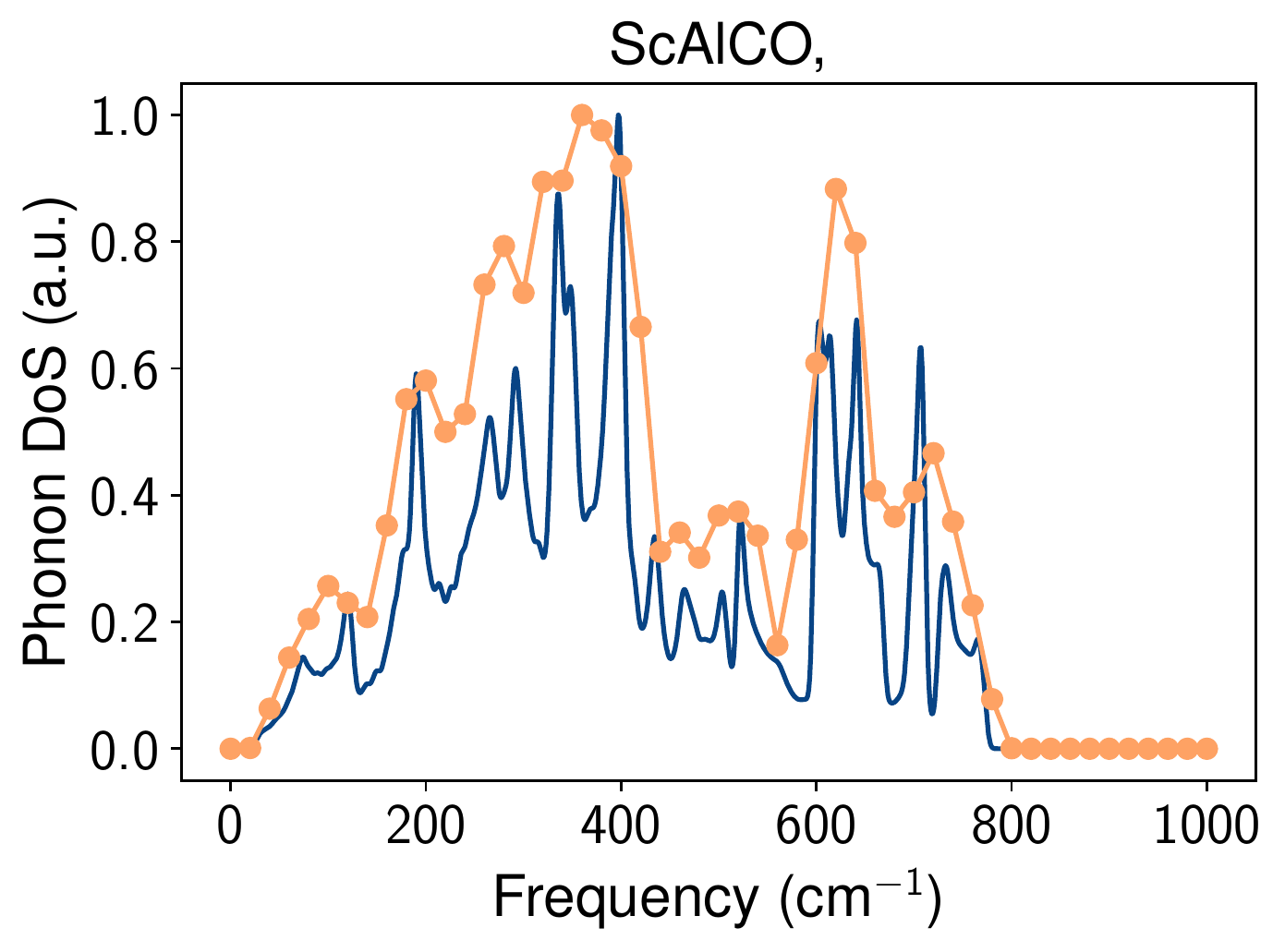}\\
    \vspace{0.5cm}
    \includegraphics[width=0.35\linewidth]{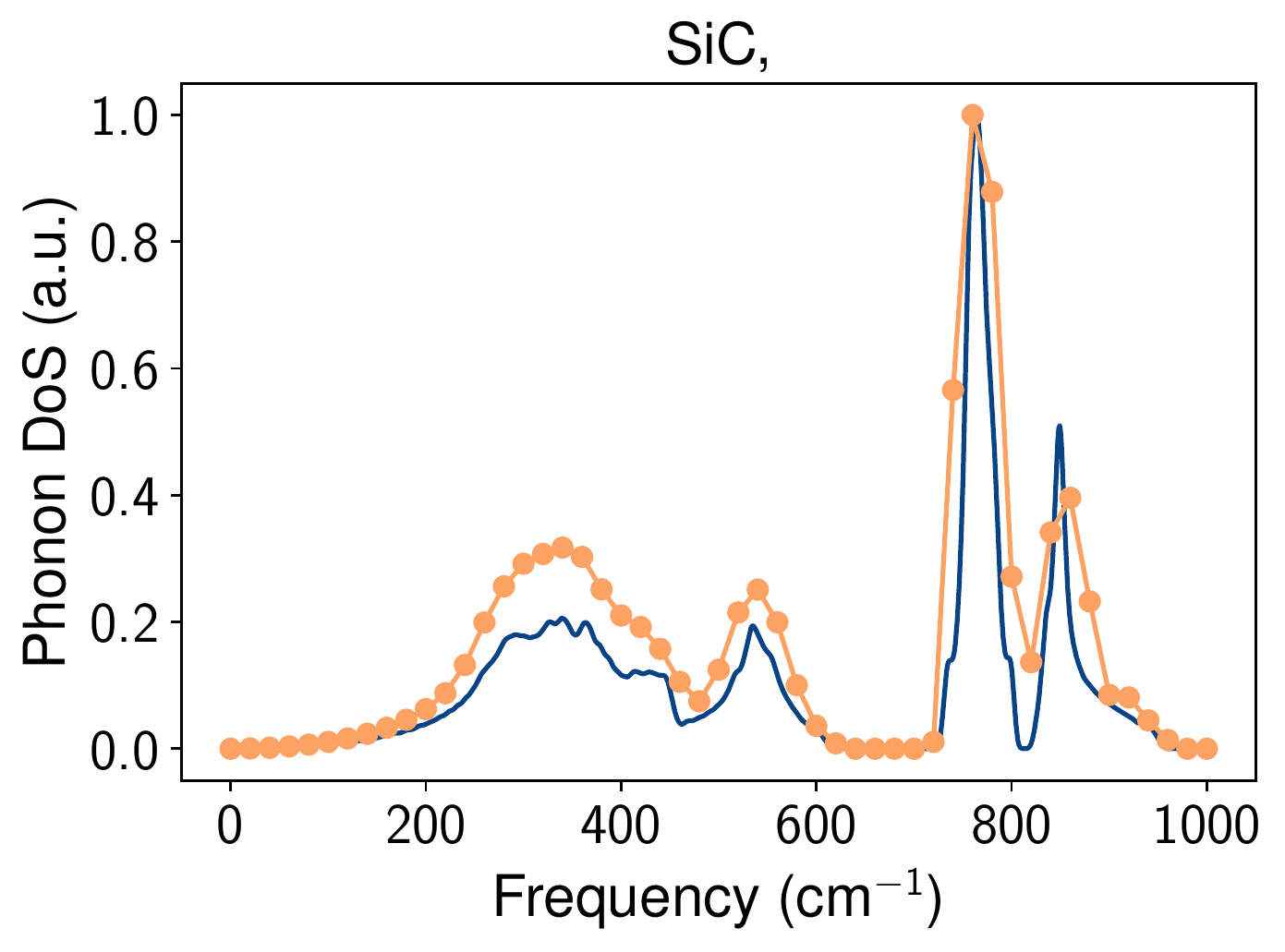}
    \hspace{1cm}
    \includegraphics[width=0.35\linewidth]{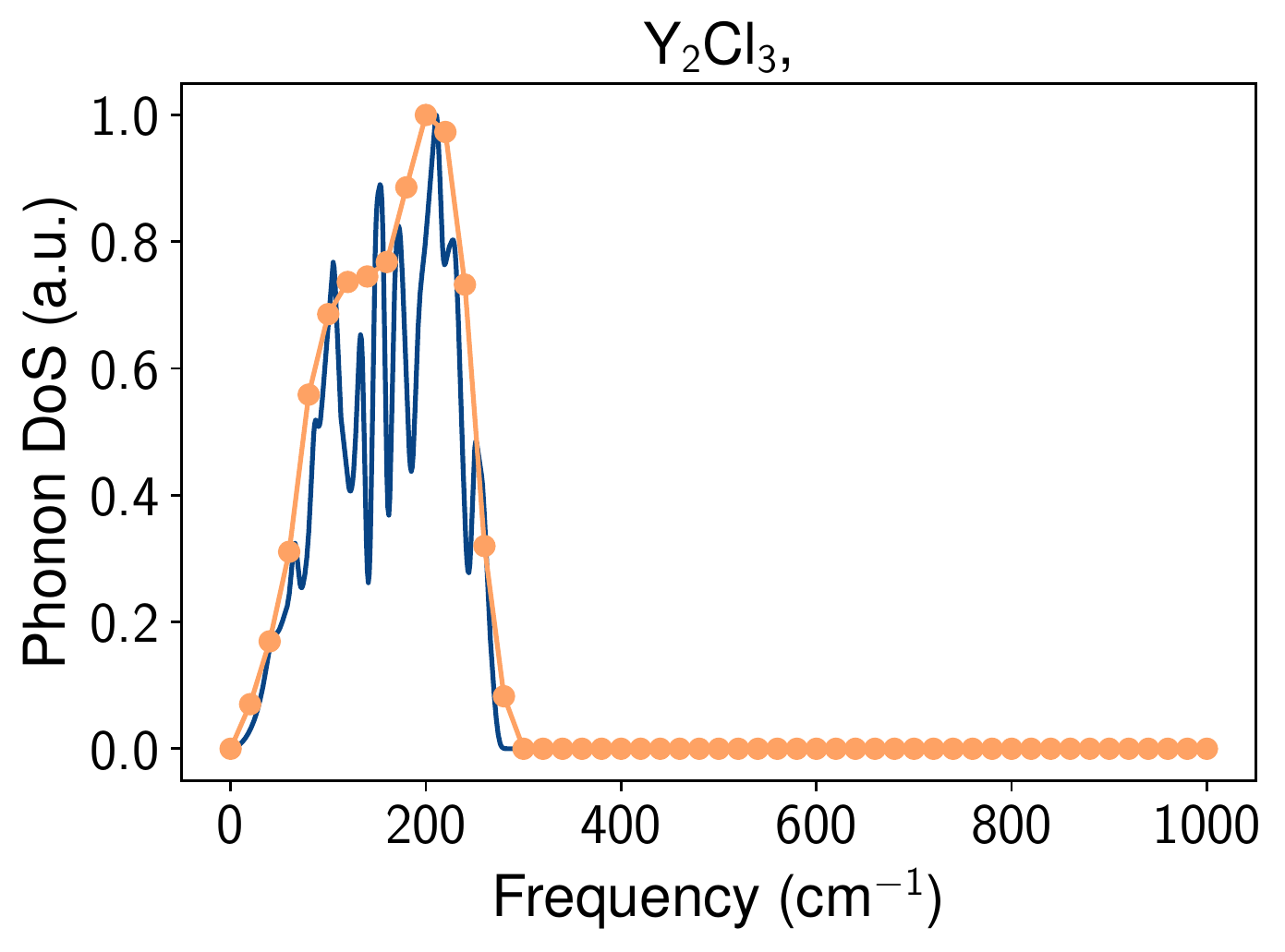}
    \caption{Original and filtered phonon DoS.}
    \label{fig:SIdosfiltering}
\end{figure}

In particular, smoothing of phonon DoS is achieved by applying a Savitzky-Golay filter of window length 101 and polynomial order 3, where filter parameters are determined to best represent raw phonon DoS: getting rid of small fluctuations while retaining the main profiles. Representative raw and smoothed phonon DoS curves are shown in Figure \ref{fig:SIdosfiltering}.

\section{Hyperparameter optimization and neural network training}

We randomly selected a set of initial hyperparameters within the ranges listed in Table \ref{tab:SIhyperparamlist} as a starting point for tuning. In addition to the listed parameters, we fixed the maximum radius $r_{\max}$ to 5\AA, which we found to adequately balance the number of edges created per vertex with the memory requirements to build the corresponding graph. The selected value is also similar to the maximum bond length considered in the CGCNN \cite{xie2018crystal}. We then systematically tested hyperparameters within the specified ranges until an optimal set balancing validation loss and memory limitations was found. The optimal set of hyperparameters, listed in the last column of Table \ref{tab:SIhyperparamlist}, was then used to train the final predictive model.

\begin{figure}
    \centering
    \includegraphics[width=0.6\linewidth]{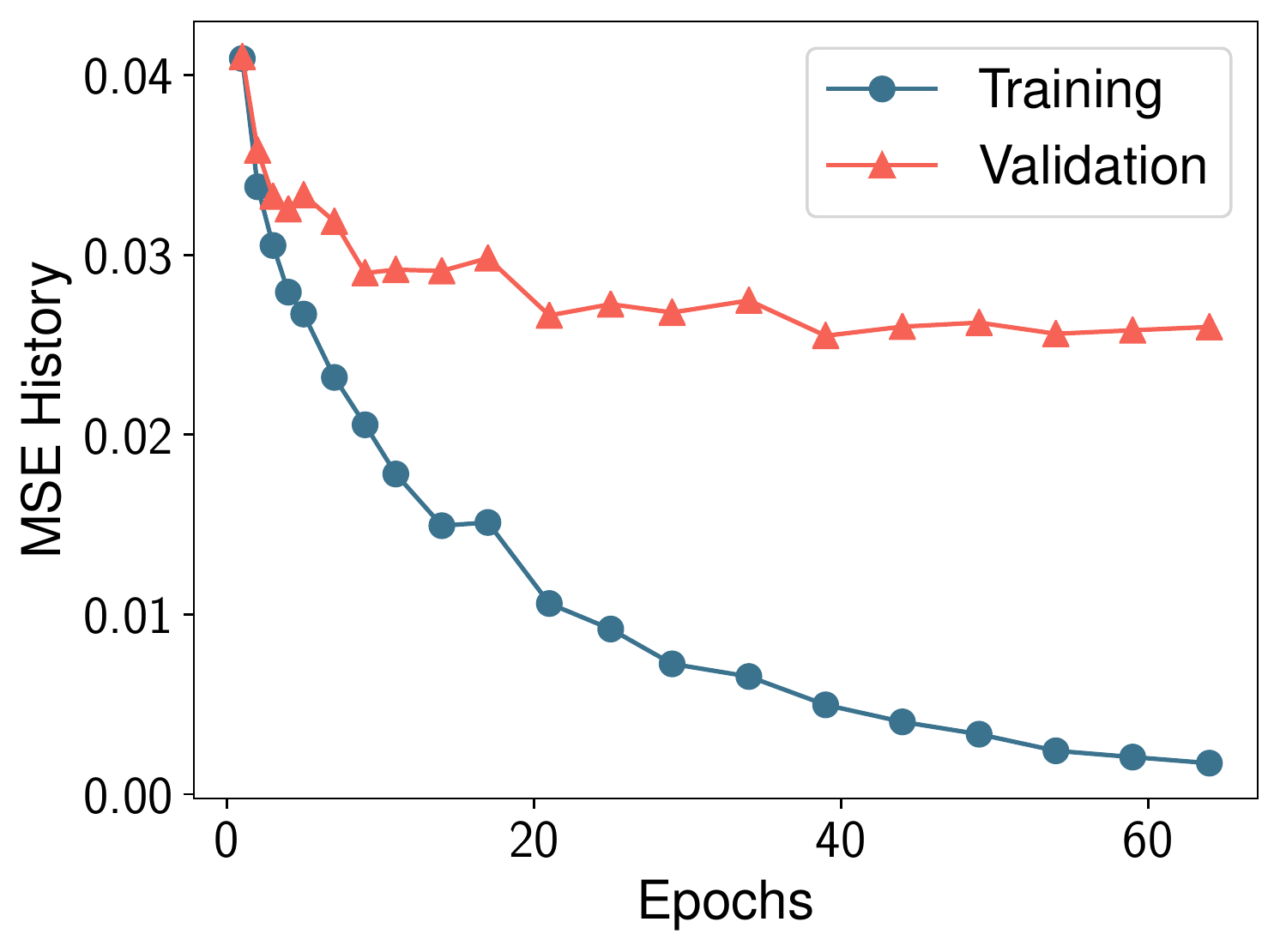}
    \caption{The mean squared error versus epoch number.}
    \label{fig:SIlosshist}
\end{figure}

The neural network is trained with a Quadro RTX 6000 GPU with 24 GB of RAM. We stop updating weights when the validation loss stagnates and tends to rise to reduce over fitting; the loss history can be found in Figure \ref{fig:SIlosshist}.

\begin{table}
	\centering
	\caption{Hyperparameter searching range and optimization process}
	\label{tab:SIhyperparamlist}
	\begin{threeparttable}
	\begin{tabular}{cccc}
		\toprule\midrule
		Hyperparameter & Range & Initial Selection & Final Selection \\
		\hline
		Multiplicity of irreducible representation $\texttt{m}$\tnote{a} & 16, 32, 48, 64\tnote{b} & 64 & 32 \\
		Number of pointwise convolution layer & 1, 2, 3, 4  & 3 & 2 \\
		Number of basis for radial filters & 5, 10, 15, 20  & 10 & 10 \\
		Length of embedding feature vector & 16, 32, 64, 128, 160  & 128 & 64 \\
		AdamW optimizer learning rate\tnote{c} & $5\mathrm{e}^{-4}$, $1\mathrm{e}^{-3}$, $5\mathrm{e}^{-3}$, $1\mathrm{e}^{-2}$  & $1\mathrm{e}^{-3}$ & $5\mathrm{e}^{-3}\times 0.96^{k}$  \\
		AdamW optimizer weight decay coefficient & $1\mathrm{e}^{-3}$, $1\mathrm{e}^{-2}$, $5\mathrm{e}^{-2}$, $1\mathrm{e}^{-1}$  & $1\mathrm{e}^{-2}$ & $5\mathrm{e}^{-2}$ \\
		\bottomrule\addlinespace[1ex]
	\end{tabular}
	\begin{tablenotes}\footnotesize
	    \item[a] For outputs of first two convolutional layers only.
		\item[b] Out of memory is encountered at value 72.
		\item[c] We observe the lowest validation loss at a learning rate of $5\mathrm{e}^{-3}$ among all values tested, followed by oscillations, and thus adopt the exponentially decaying learning rate with $k$ being the epoch number.
	\end{tablenotes}
	\end{threeparttable}
\end{table}

\newpage
{\color{revision}
\section{Model limitations on strained samples}
\label{sec:SIstrain}

The proposed machine learning approach has demonstrated success in unseen elements and alloys, but faces one limitation on strained sample at this moment. In Figure \ref{fig:SI_STO_strain}, we compare the neural network predictions with DFPT calculations for SrTiO\textsubscript{3} at three levels of strains. In both cases, the investigated supercells can be evaluated within seconds by a trained E(3)NN neural network. As we can see that the DFPT-computed results are more sensitive than those from E(3)NN, which is attributed to a lack of knowledge of equilibrium bond lengths to be used for the model training and the limited coverage of training data, for example, confined energy range $0\le \omega\le 1000\, \mathrm{cm}^{-1}$ without consideration of imaginary modes. However, better performances are still feasible if more relevant strain-dependent training data can be provided.

\begin{figure}
    \centering
    \includegraphics[width=0.6\linewidth]{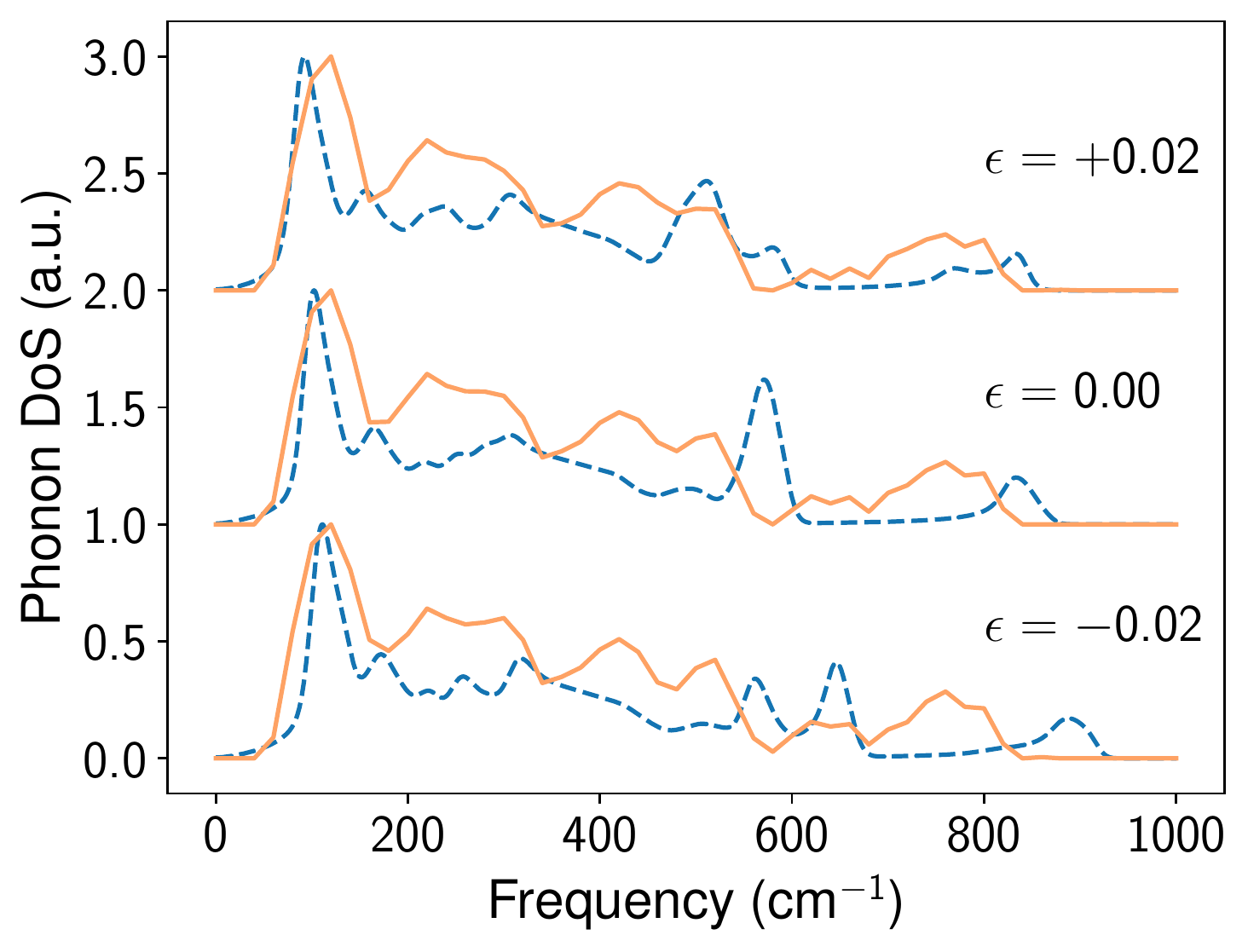}
    \caption{Phonon DoS of SrTiO\textsubscript{3} with biaxial strain, including compressional (upper), unstrained (middle), and tensile (lower). The blue-dashed lines and orange lines represent DFPT and E(3)NN predictions, respectively.}
    \label{fig:SI_STO_strain}
\end{figure}

}

\clearpage
{\color{revision2}
\section{Training with imaginary phonon mode}

Considering the fact that in the dataset \cite{petretto2018high} around 1/6 materials have relatively large negative frequencies in their calculated phonon DoS, indicating a possible poor convergence in the DFPT calculation. We plot the average phonon DoS in Figure \ref{fig:SI_avg_phonondos}. The majority of calculated materials have insignificant imaginary modes, with a mean phonon frequency of -6.11 cm$^{-1}$ and a STD of 16.91 cm$^{-1}$ (computed from uninterpolated original data in \cite{petretto2018high}), computed with the same equation $\bar{\omega} = \tfrac{\int \mathrm{d}\omega \, g(\omega) \, \omega }{\int \mathrm{d}\omega \, g(\omega)}$ used in Figure \ref{fig:trainperformance}. It is noticed that most negative frequency phonon modes are located above $-200\,\mathrm{cm}^{-1}$. In order to evaluate the influences of including imaginary phonon modes, we expand the frequency range to $-200\le \omega\le 1000\,\mathrm{cm}^{-1}$ for our model, with same energy interval $20\,\mathrm{cm}^{-1}$ between sampled points such that the final output is an array of length 61, and keep all other hyperparameters unchanged.

In Figure \ref{fig:SI_loss_imagmode}, we visualize the losses of the model trained with imaginary phonon modes, where very similar loss distributions are obtained compared with the model trained without imaginary modes in the main article (Figure \ref{fig:trainperformance}). We further plot true (DFPT) and predicted phonon DoS for materials from training, validation, and test sets in Figure \ref{fig:SI_imagmode_tr}, \ref{fig:SI_imagmode_va}, and \ref{fig:SI_imagmode_te}. The example materials in each figure are ordered by the mean magnitude of phonon DoS at negative frequencies. We find that for many materials in the test set, the predictions made in positive energy range are significantly better than in negative frequency ranges, such as Rb\textsubscript{2}TlInF\textsubscript{6}, RbHgF\textsubscript{3}, Cs\textsubscript{2}As\textsubscript{3}, etc. Similar trends can also be observed for the other two data sets. On one hand, the poor performance on the negative frequencies could come from much less informative training samples. On the other hand, the performance on the positive frequencies of our model seems not to be affected significantly by those information.

\begin{figure}[h]
    \centering
    \includegraphics[width=0.8\linewidth]{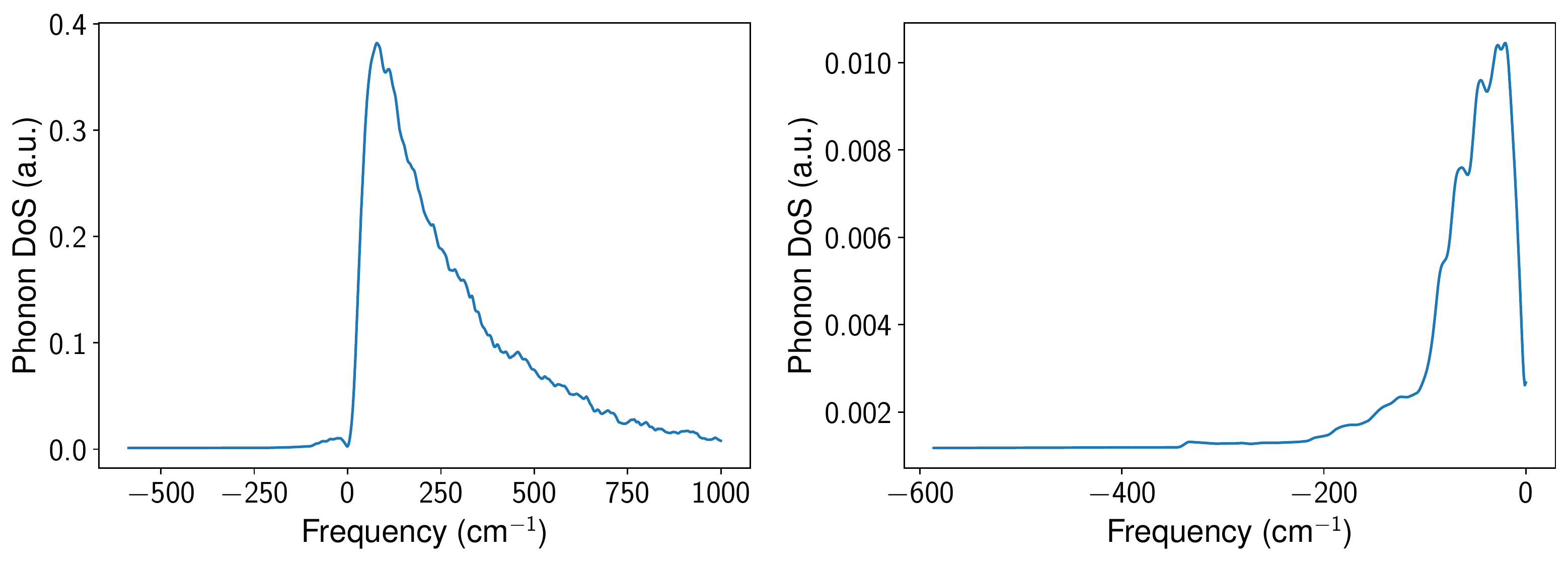}
    \caption{Mean phonon DoS averaged from all 1,521 examples in \cite{petretto2018high}.}
    \label{fig:SI_avg_phonondos}
\end{figure}

\begin{figure}[h]
    \centering
    \includegraphics[width=0.35\linewidth]{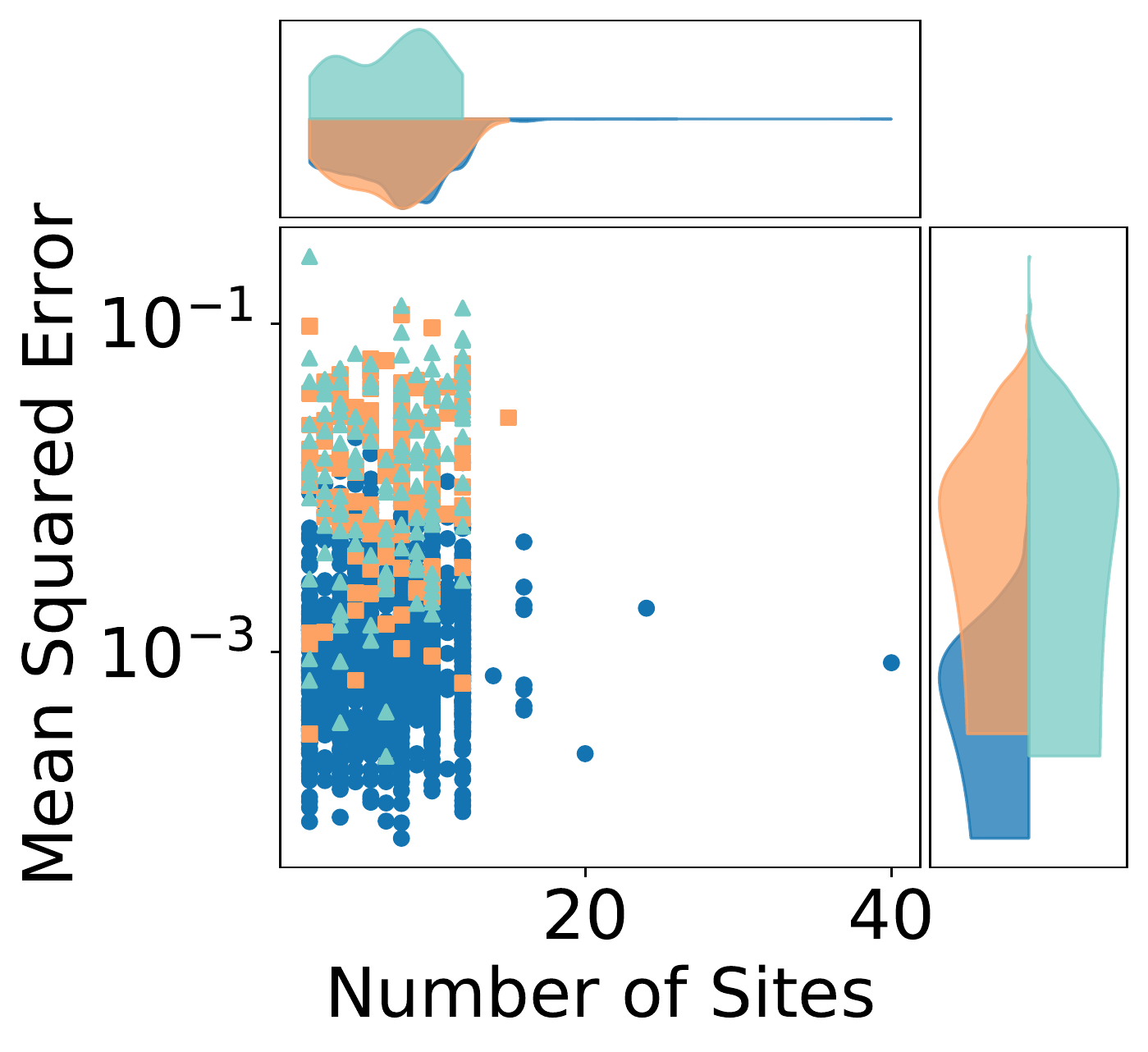}
    \hspace{10pt}
    \includegraphics[width=0.35\linewidth]{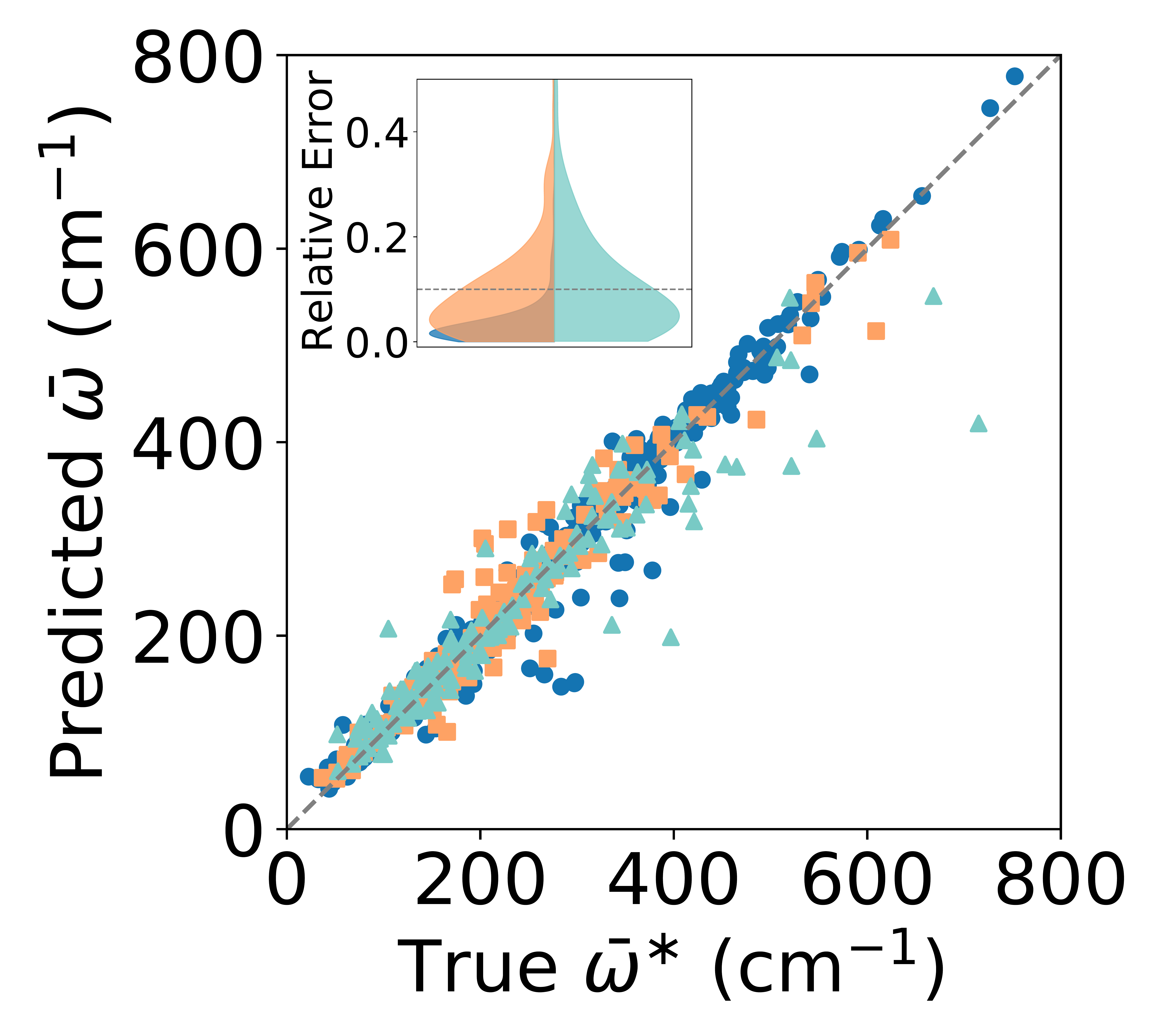}
    \caption{Loss visualization for the model trained with imaginary phonon modes.}
    \label{fig:SI_loss_imagmode}
\end{figure}

\begin{figure}
    \centering
    \includegraphics[width=0.7\linewidth]{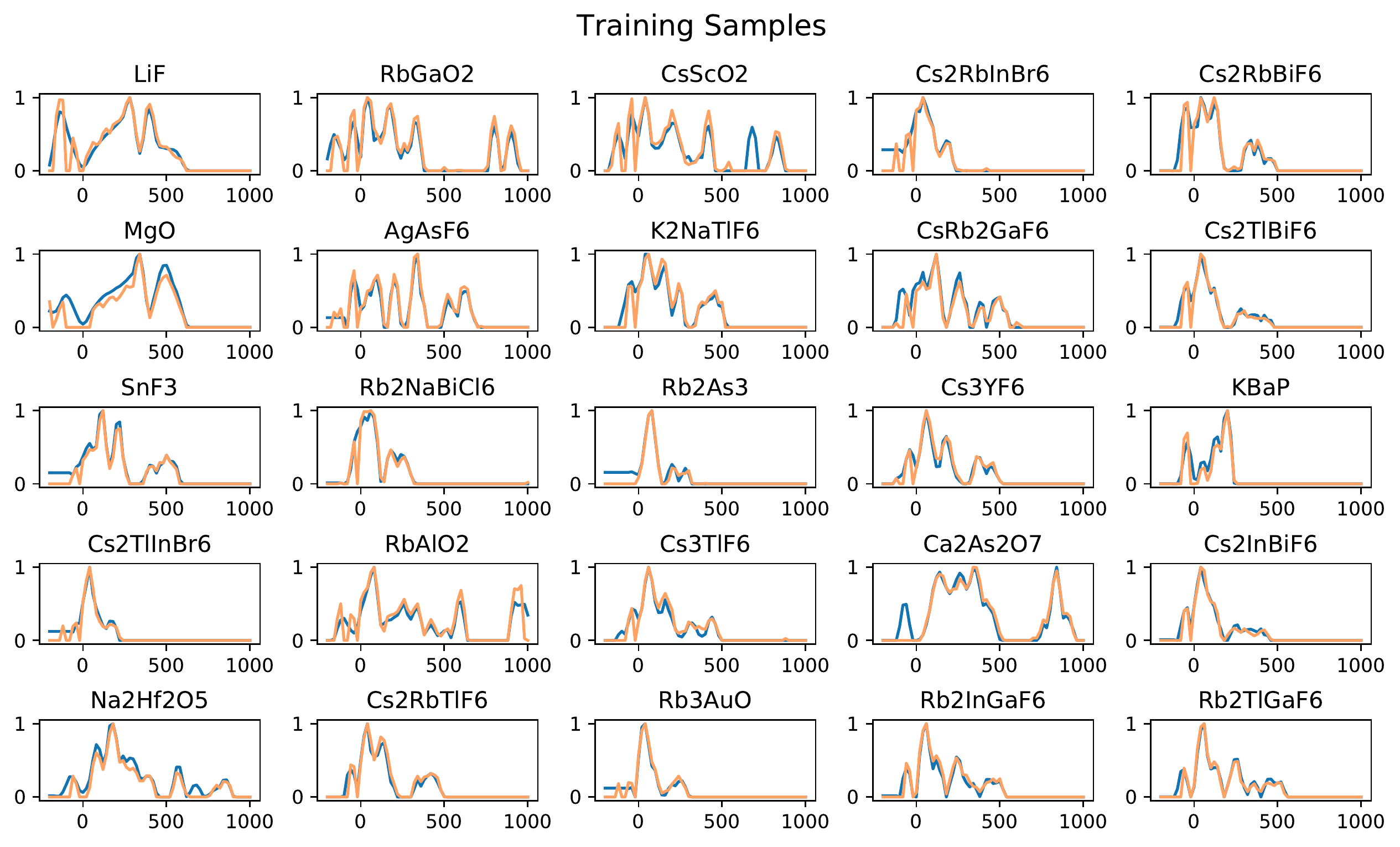}
    \caption{Predicted phonon DoS within training set. Blue and orange curves represent true (DFPT) and predicted phonon DoS, respectively.}
    \label{fig:SI_imagmode_tr}
\end{figure}

\begin{figure}
    \centering
    \includegraphics[width=0.7\linewidth]{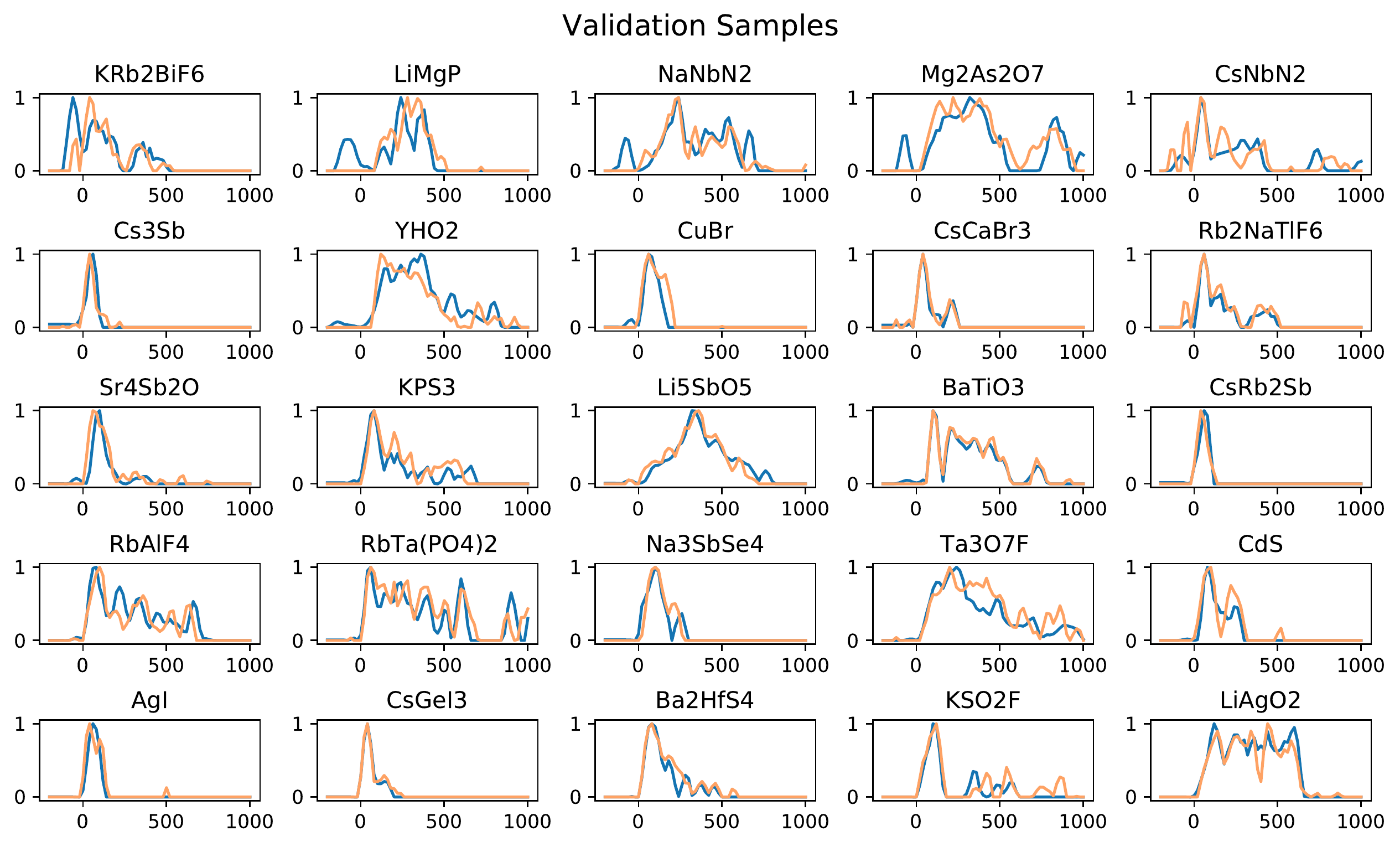}
    \caption{Predicted phonon DoS within validation set. Blue and orange curves represent true (DFPT) and predicted phonon DoS, respectively.}
    \label{fig:SI_imagmode_va}
\end{figure}

\begin{figure}
    \centering
    \includegraphics[width=0.7\linewidth]{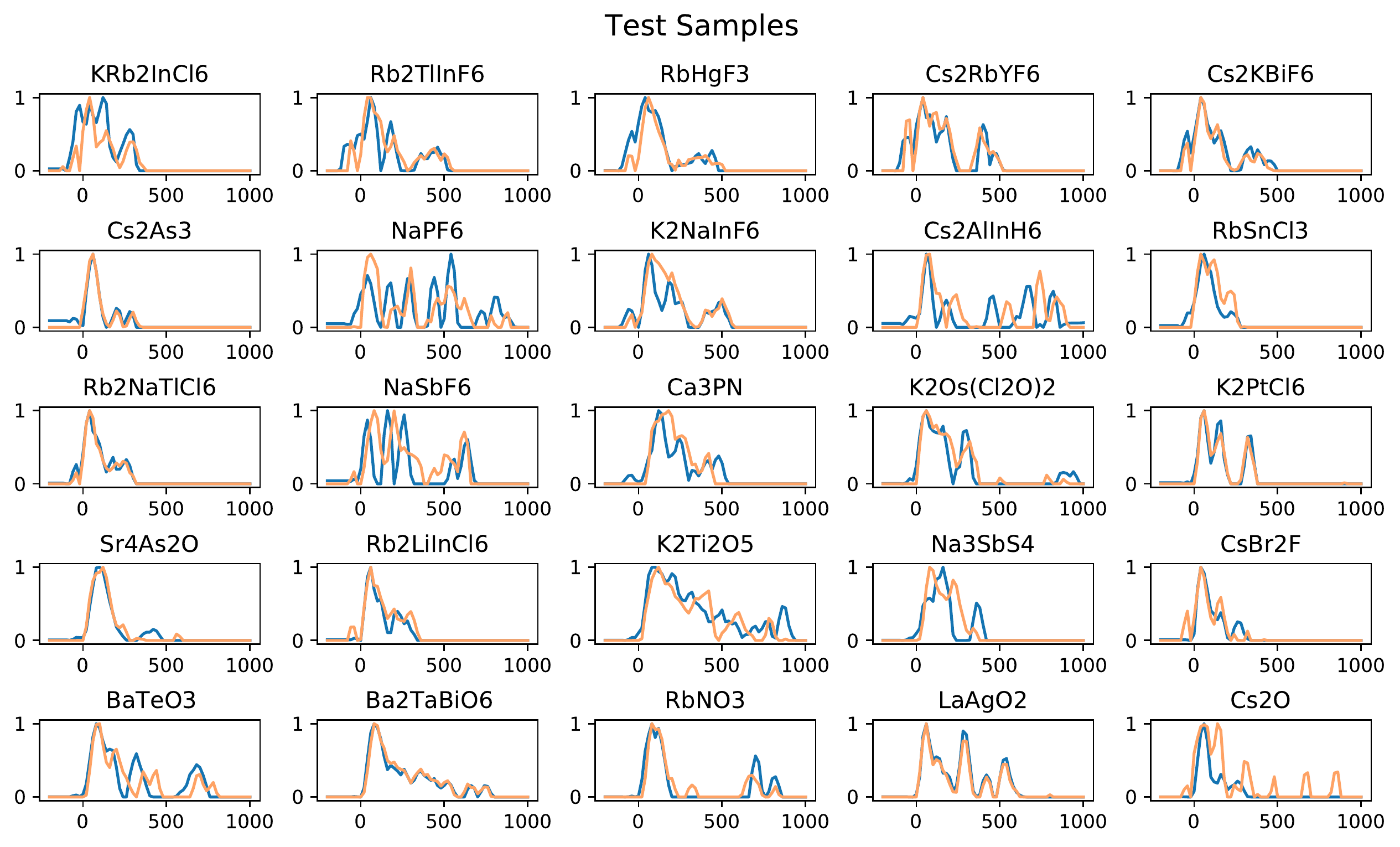}
    \caption{Predicted phonon DoS within test set. Blue and orange curves represent true (DFPT) and predicted phonon DoS, respectively.}
    \label{fig:SI_imagmode_te}
\end{figure}
}
\clearpage
\section{More predicted phonon DoS in test set}
\label{sec:morepredictedphonondos}

\begin{figure}[h]
    \centering
    \includegraphics[width=0.7\linewidth]{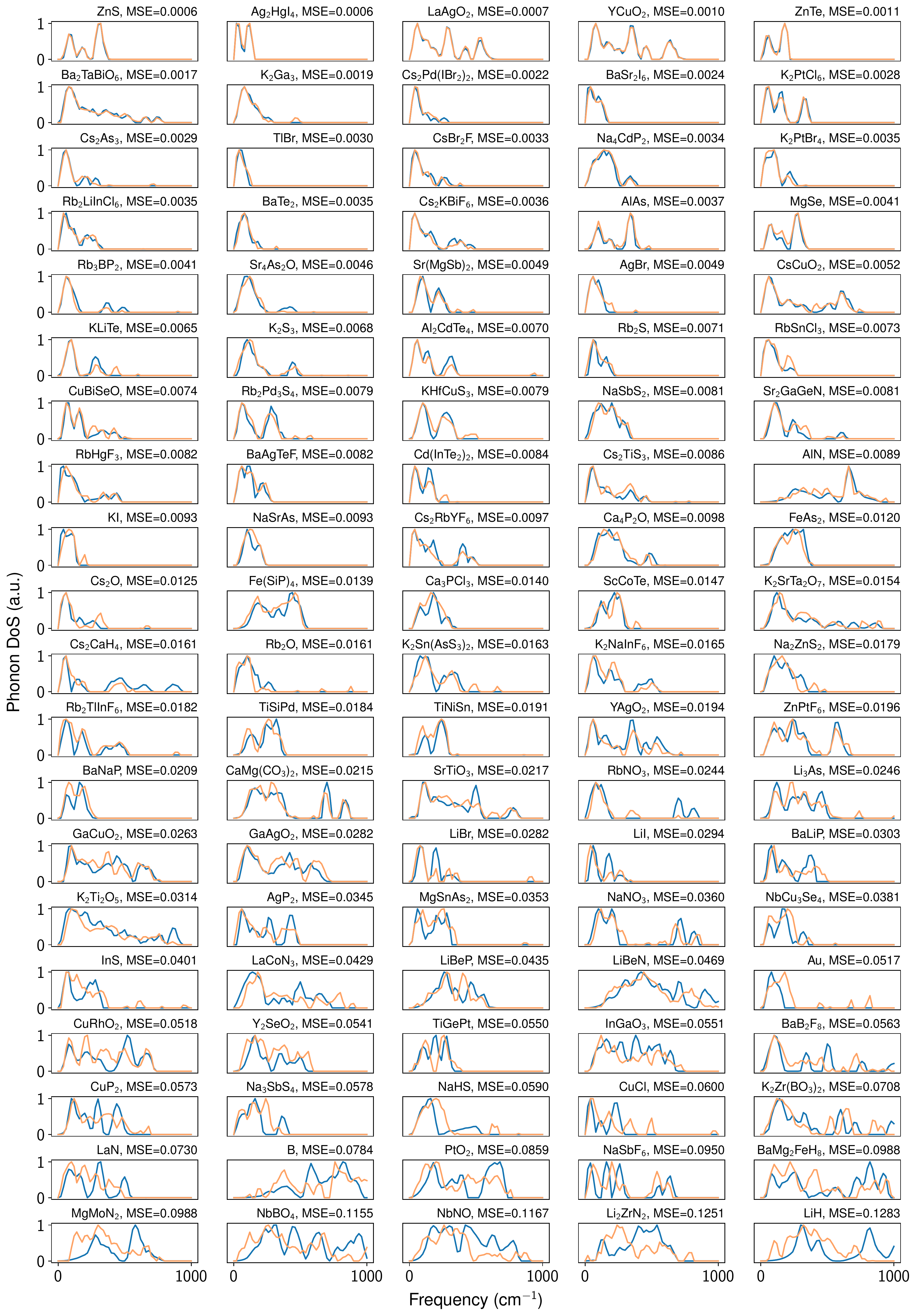}
    \caption{Predicted phonon DoS of randomly-selected examples from the test set, sorted from top to bottom in order of increasing MSE.}
    \label{fig:SImoreTestDoS}
\end{figure}

We present 100 predicted phonon DoS from the test set on opt of those appeared in the main text in Figure \ref{fig:SImoreTestDoS}. The samples are sorted from top to bottom in order of increasing mean squared error (MSE). As discussed in the main text, our model performs better in predicting energy ranges than exact amplitudes. While near-perfect predictions were achieved for examples in the first few rows, the principal peaks and gaps of the phonon DoS were well-reproduced for many materials located at bottom rows, such as MgSnAs\textsubscript{2}, LiBeP, InGaO\textsubscript{3}, and NaSbF\textsubscript{6}.

\newpage
\section{Element representation}

\begin{figure}[h]
    \centering
    \includegraphics[width=0.7\linewidth]{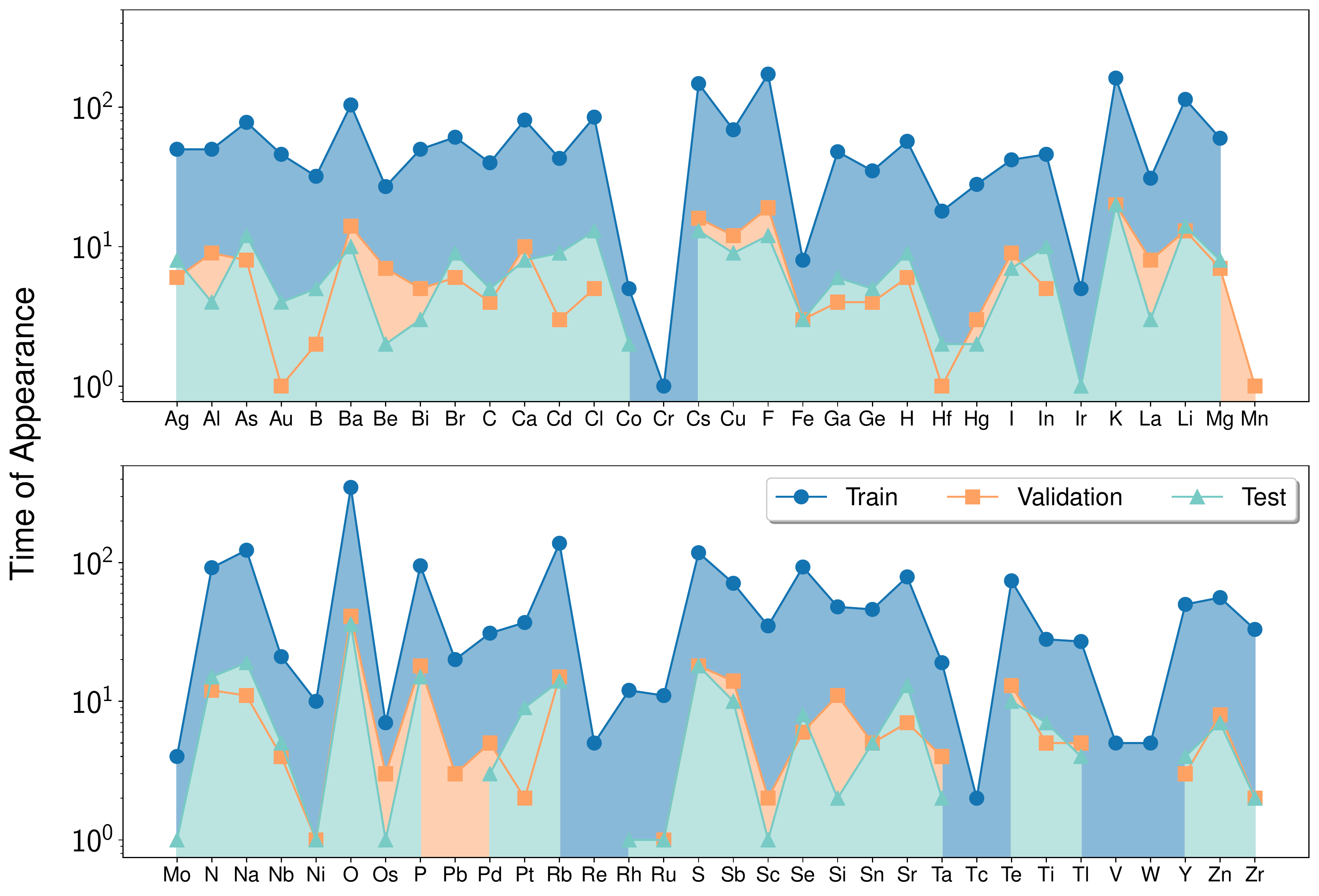}
    \caption{Element representations in training, validation, and test sets.}
    \label{fig:SIspeciescount}
\end{figure}

\begin{figure}[h]
    \centering
    \includegraphics[width=0.7\linewidth]{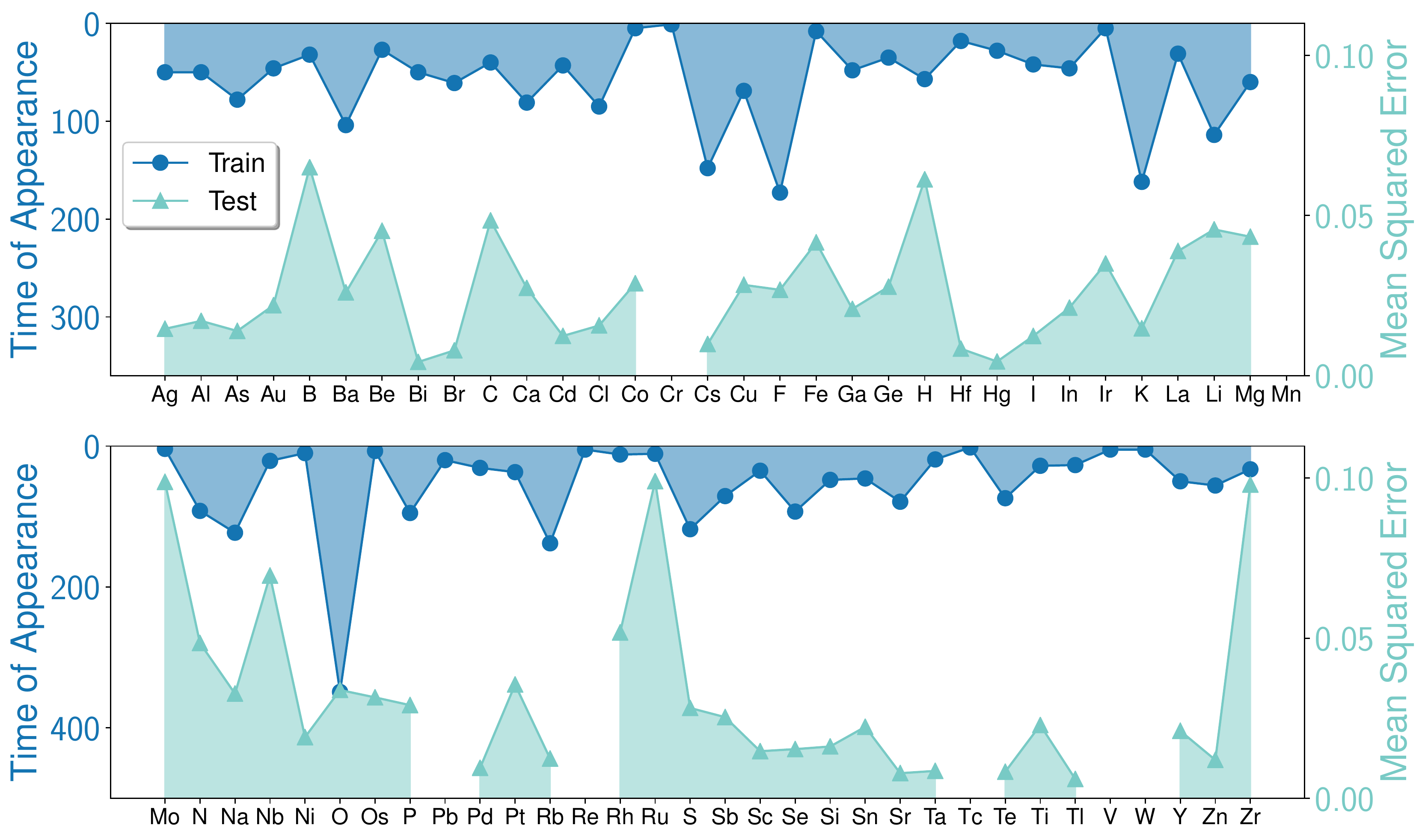}
    \caption{Element representation in training set versus MSE in test set. The inset shows a scatter plot of the relationship.}
    \label{fig:SIAPPTIMEvsMSE}
\end{figure}

In Figure \ref{fig:SIspeciescount}, we illustrate the number of occurrences of each element in the training, validation, and test sets. When compared to Figure \ref{fig:trainperformance}b in the main text, it is found that elements corresponding to high MSE in Figure 2b of the main text tend to appear less frequently than those with lower errors, such as La, Ir, Mo, Rh, and Ru. A more direct comparison is made in Figure \ref{fig:SIAPPTIMEvsMSE}.

\newpage
\section{Additional phonon DoS predictions of materials with highest specific heat capacity}

\begin{figure}[h]
    \centering
    \includegraphics[width=0.7\linewidth]{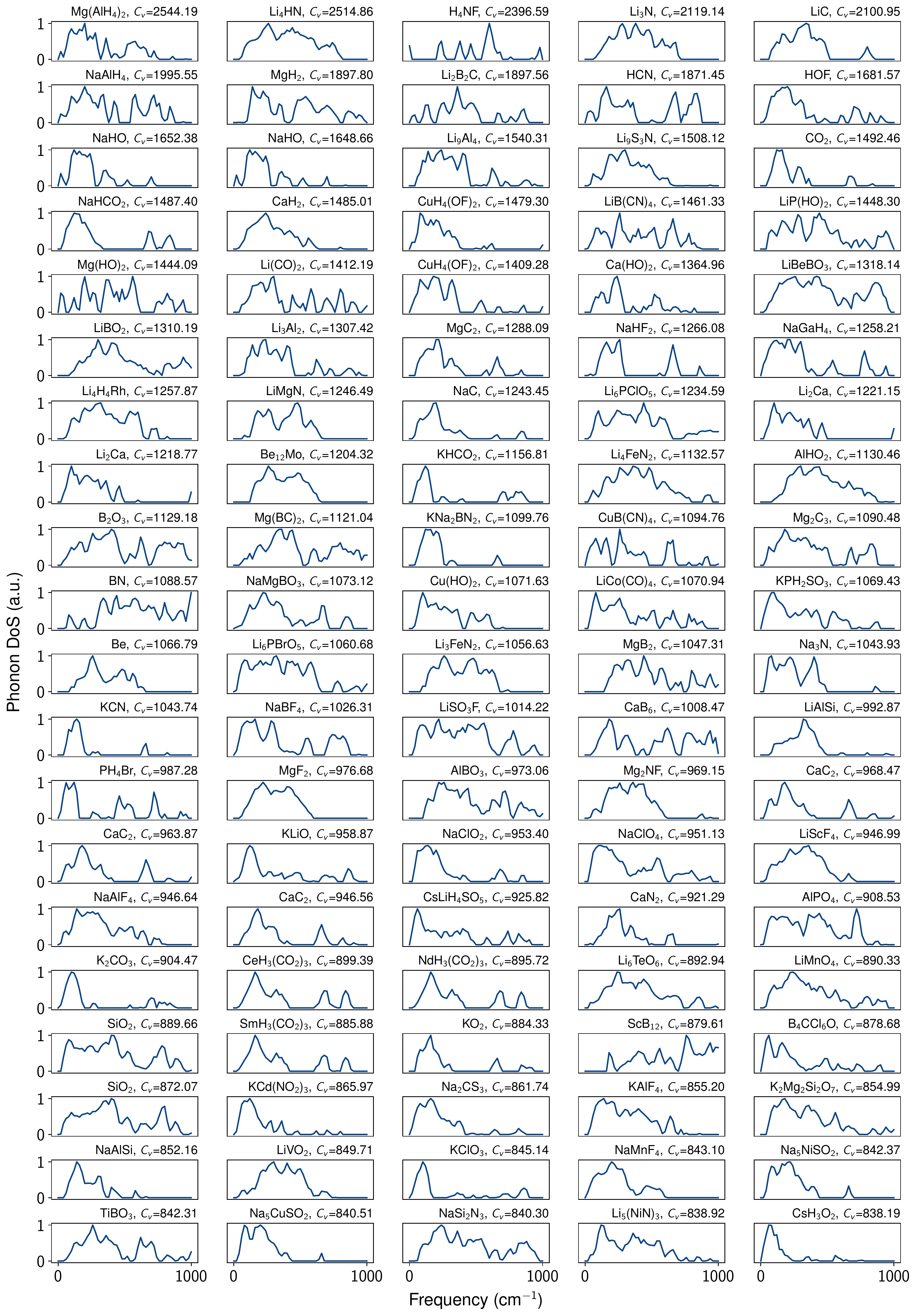}
    \caption{Top 100 high specific heat capacity materials, $C_{V}$ in unit $\mathrm{J/(kg\cdot K)}$.}
    \label{fig:SItop100Cv}
\end{figure}

We present E(3)NN-predicted materials (selected from 4,346 crystal structures without ground-truth DoS from the Materials Project \cite{jain2013commentary}) with the top 100 highest specific heat capacities in Figure \ref{fig:SItop100Cv}.

\newpage
\section{Specific heat capacity calculated from full energy ranges}
\label{sec:fullrangedos}

Due to considerations about neural network size and spatial resolution of the predicted phonon DoS, our model is trained with DFPT-calculated phonon DoS in the energy range $[0,1000]\, \mathrm{cm^{-1}}$. When evaluating new materials, the E(3)NN-predicted phonon DoS is also confined within this domain. Therefore the specific heat capacities are evaluated from potentially incomplete phonon DoS. 

\begin{figure}[h]
    \centering
    \includegraphics[height=0.4\linewidth]{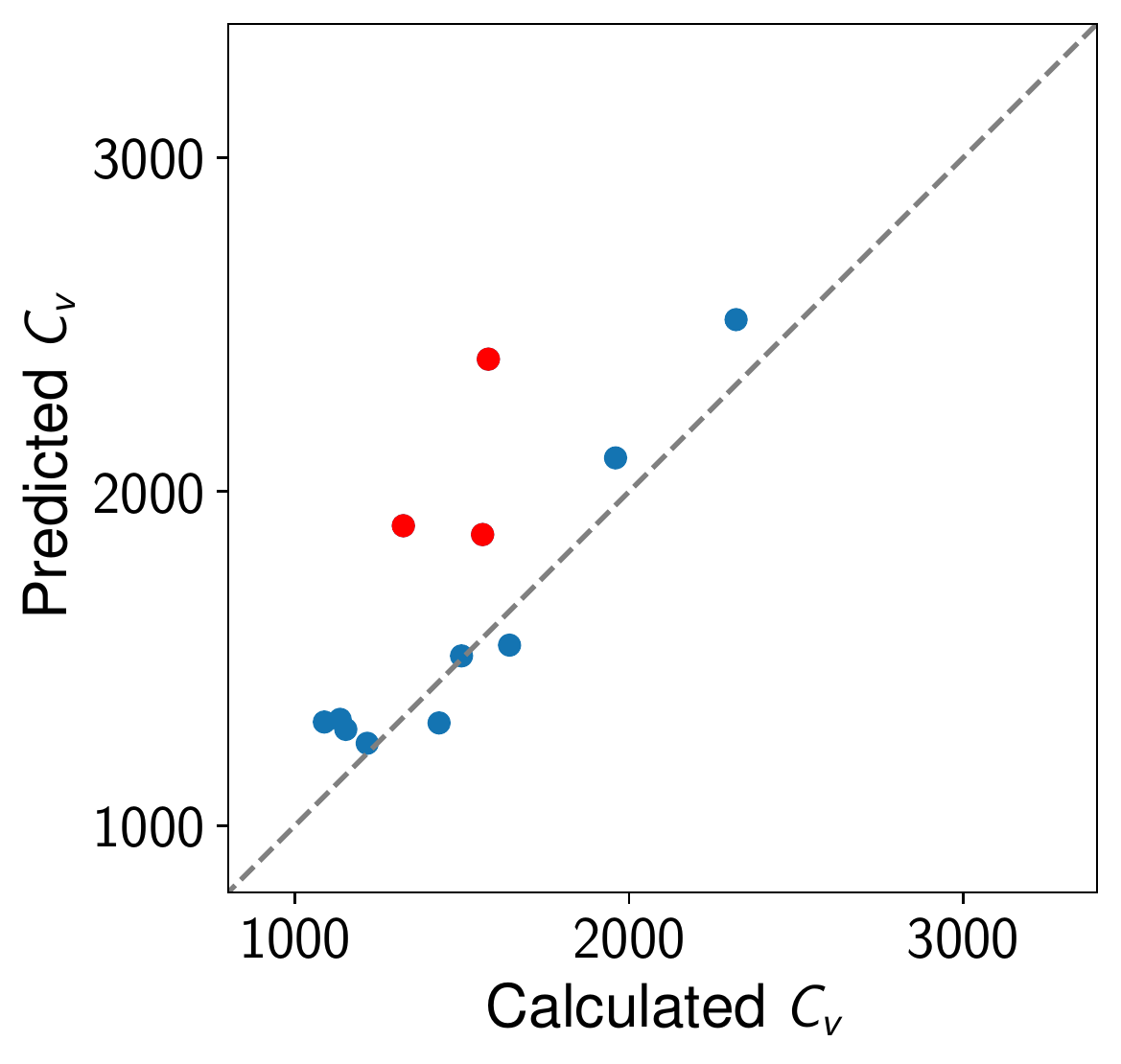}
    \hspace{1cm}
    \includegraphics[height=0.4\linewidth]{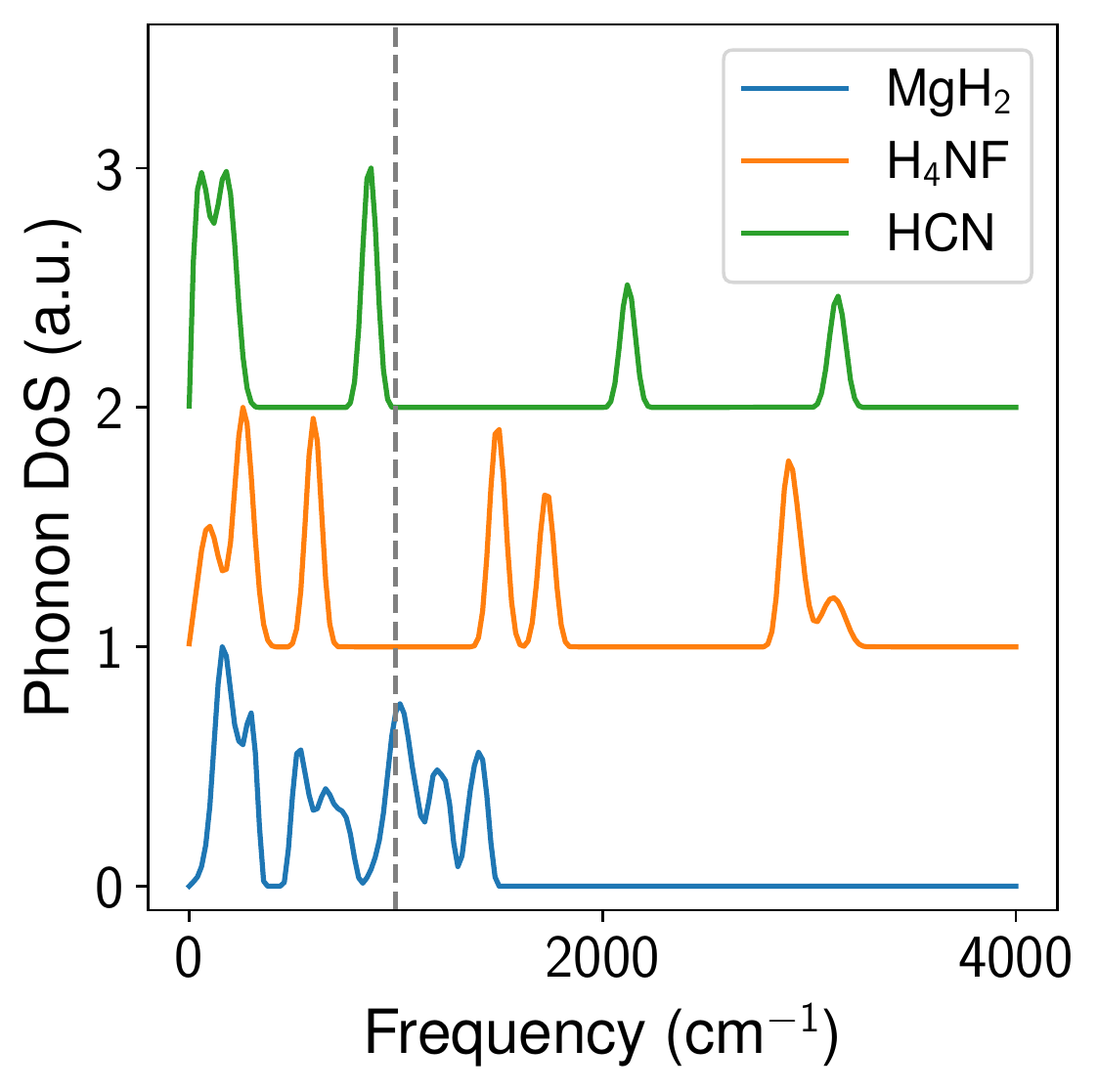}
    \caption{Left: Two-dimensional histogram comparing between specific heat capacities evaluated from E(3)NN-predicted and complete DFPT-calculated phonon DoS. Right: Full phonon DoS of three materials that display largest discrepancies.}
    \label{fig:SIdoscutoff}
\end{figure}

Here we present the specific heat capacities evaluated from the confined and complete phonon DoS in Table \ref{tab:SIcompareheatcap}. It is found that most materials have a slightly different $C_{V}$, and 9 out of 12 points are still in good agreement. On the other hand, more dramatic discrepancies in the other 3 materials, MgH\textsubscript{2}, H\textsubscript{4}NF, and HCN, are partially induced by significant contributions to the DoS at higher energy ranges outside those considered during training. The two-dimensional histogram and corresponding phonon DoS for the three materials are shown in Figure \ref{fig:SIdoscutoff}. A potential remedy for this issue is to train our model with phonon DoS including higher energy ranges which may lose the energy resolution and is not necessary of majority of materials. The good agreement between E(3)NN-predictions and DFPT ground-truth spectra achieved in the energy range chosen for this work sufficiently demonstrates the potential for generalizability of our model.

\begin{table}[h]
    \centering
    \caption{Comparisons of E(3)NN predicted and DFPT calculated specific heat capacities (in unit $\mathrm{J /(kg\cdot K)}$). Left: $C_{V}$ is computed from energy range $0\le \omega\le 1000\,\mathrm{cm^{-1}}$. Right: $C_{V}$ is computed from full energy range.}
    \vspace{5pt}
    \begin{tabular}{ccccccc}
    \hline\hline
    Material & $C_V^{\text{E(3)NN}}$ & $C_V^{\text{DFPT}}$ & \hspace{0.25cm} & Material & $C_V^{\text{E(3)NN}}$ & $C_V^{\text{DFPT}}$ \\
    \hline
    Li\textsubscript{4}HN  & 2514.9 & 2320.2 & & Li\textsubscript{9}S\textsubscript{3}N & 1508.1 & 1498.3\\
    H\textsubscript{4}NF & 2396.6 & 3117.4 & & LiBeBO\textsubscript{3} & 1318.1 & 1286.1\\
    LiC & 2101.0 & 2138.8 & & LiBO\textsubscript{2} & 1310.2 & 1172.9\\
    MgH\textsubscript{2} & 1897.8 & 1908.5 & & Li\textsubscript{3}Al\textsubscript{2} & 1307.4 & 1431.2\\
    HCN & 1871.5 & 2047.2 & & MgC\textsubscript{2} & 1288.1 & 1307.8\\
    Li\textsubscript{9}Al\textsubscript{4} & 1540.3 & 1642.1 & & LiMgN & 1246.5 & 1216.2\\
    \hline\hline
    \end{tabular}
    \hspace{.5cm}
    \begin{tabular}{ccccccc}
    \hline\hline
    Material & $C_{V}^{\text{E(3)NN}}$ & $C_{V}^{\text{DFPT}}$ & \hspace{0.25cm} & Material & $C_{V}^{\text{E(3)NN}}$ & $C_{V}^{\text{DFPT}}$ \\
    \hline
    Li\textsubscript{4}HN  & 2514.86 & 2320.06 & & Li\textsubscript{9}S\textsubscript{3}N & 1508.12 & 1498.39\\
    H\textsubscript{4}NF & 2396.59 & 1578.65 & & LiBeBO\textsubscript{3} & 1318.14 & 1134.99\\
    LiC & 2100.95 & 1959.36 & & LiBO\textsubscript{2} & 1310.19 & 1087.88\\
    MgH\textsubscript{2} & 1897.80 & 1324.31 & & Li\textsubscript{3}Al\textsubscript{2} & 1307.42 & 1431.15\\
    HCN & 1871.45 & 1561.42 & & MgC\textsubscript{2} & 1288.09 & 1152.05\\
    Li\textsubscript{9}Al\textsubscript{4} & 1540.31 & 1642.07 & & LiMgN & 1246.49 & 1216.17\\
    \hline\hline
    \end{tabular}
    \label{tab:SIcompareheatcap}
\end{table}


\section{Comparison to convolutional neural network approach}

\begin{table}
\centering
\caption{Architectures of the convolutional neural network.}
\label{tab:SIconvneuralnet}
\begin{tabular}{lcc}
\toprule\midrule
Layer                                                     & Input shape                   & Output shape                \\ \hline
3D conv. layer, kernel size=5, strides=2, channel num.=1 & \multirow{2}{*}{(32,32,32,2)} & \multirow{2}{*}{(16,16,16,1)} \\ \cline{1-1}
activation=ReLU                                           &                               &                             \\ \hline
3D conv. layer, kernel size=3, strides=2, channel num.=2  & \multirow{2}{*}{(16,16,16,1)} & \multirow{2}{*}{(8,8,8,2)}  \\ \cline{1-1}
activation=ReLU                                           &                               &                             \\ \hline
3D conv. layer, kernel size=3, strides=2, channel num.=4 & \multirow{2}{*}{(8,8,8,2)}    & \multirow{2}{*}{(4,4,4,4)} \\ \cline{1-1}
activation=ReLU                                           &                               &                             \\ \hline
3D conv. layer, kernel size=3, strides=2, channel num.=8 & \multirow{2}{*}{(4,4,4,4)}   & \multirow{2}{*}{(2,2,2,8)} \\ \cline{1-1}
activation=ReLU                                           &                               &                             \\ \hline
Flatten                                                   & (2,2,2,8)                    & 64                         \\
Fully connected, activation=ReLU                          & 64                           & 64                          \\
Fully connected, activation=ReLU                          & 64                            & 51                          \\ 
\bottomrule
\end{tabular}
\end{table}

\begin{figure}
    \centering
    \includegraphics[width=0.9\linewidth]{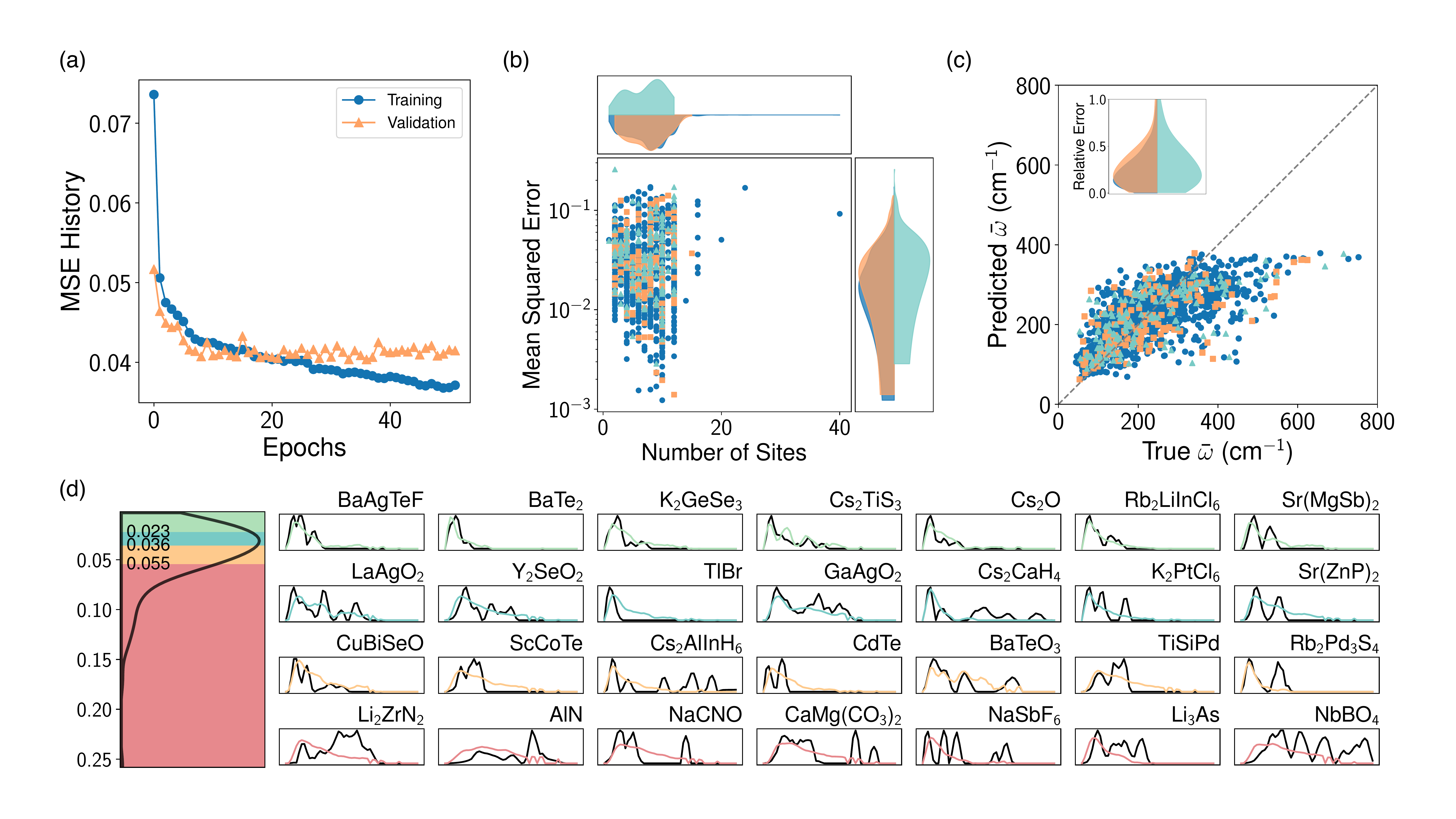}
    \caption{Phonon DoS predicted by the convolutional neural network.}
    \label{fig:SI_CNN_model_performance}
\end{figure}

In order to better quantify the advantages of the tensor field neural network, we present a brief comparison to a convolutional neural network (CNN). Here, the unit cells are encoded as three-dimensional densities, similar to \cite{hoffmann2019data}. In particular, they are defined by 
\begin{equation}
    \rho_{Z}(\bm{r})=\sum_{i=1}^{N}Z_{i}\mathrm{e}^{-|\bm{r}-\bm{r}_{i}|^{2}},\qquad \rho_{m}(\bm{r})=\sum_{i=1}^{N}m_{i}\mathrm{e}^{-|\bm{r}-\bm{r}_{i}|^{2}},
\end{equation}
and mapped onto a discrete three-dimensional grid with positions of grid points determined by corresponding lattice vectors and parameters. The densities are then input into a three-dimensional convolutional neural network (architecture is shown in Table \ref{tab:SIconvneuralnet}). Same training, validation, and test samples are used here. In particular, to consider lattice periodicity, the densities for each unit cell are calculated in the middle of a $3\times3\times3$ super cell to account for contributions from neighbors. For each material, four unit cells with randomly translated atomic positions (shifted origins) alongside the original unit cell are used to train and validate the model, in order to achieve representation-independent outputs. The resulting performances are summarized in Figure \ref{fig:SI_CNN_model_performance} and are found not as good as E(3)NN's. Specifically, the CNN seems intended to predict similar profiles regardless of specific atomic structures, and unable to capture high energy peaks. Such performance further results in lower predicted average frequency, the failure in capturing main features makes it unsuitable to be applied on unseen materials.

Admittedly, there are several ways to improve the current CNN architecture. One noticeable problem is that the convolutional kernel here acts on different length scales depending on the lattice. A fixed grid like in \cite{hoffmann2019data} can also be adopted to resolve this issue, but other constraints such as limited representation ability (confined by grid dimension) and lowered spatial resolution will be brought up. In addition, more data augmentation will also be helpful. However, the E(3)NN offered a more physics-informed framework to intrinsically consider rotation and translation equivariance while get rid of data augmentation to capture arbitrary orientations and periodicity of the unit cell. In this case, the parameters in E(3)NN are more focused on building direct connections between crystal structures and target properties without interference from other information. As a result, it can achieve higher prediction accuracy and better generalization ability.



\end{document}